\documentclass[review]{elsarticle}

\usepackage{array}
\usepackage{color}
\usepackage{fancyhdr}
\usepackage{geometry}
\usepackage{graphicx}
\usepackage{hyperref}
\usepackage{multirow}
\usepackage[table]{xcolor}
\usepackage{amsmath}
\usepackage{comment}

\usepackage{float}
\usepackage{amssymb}

\usepackage[ruled]{algorithm2e} 
\usepackage{color}

\def\HiLiDR{\leavevmode\rlap{\hbox to \hsize{\color{red!20}\leaders\hrule height .8\baselineskip depth .5ex\hfill}}}
\def\HiLiCL{\leavevmode\rlap{\hbox to \hsize{\color{green!10}\leaders\hrule height .8\baselineskip depth .5ex\hfill}}}

\begin{document}
\begin{frontmatter}

%
\title{A Distributed Learning Architecture for Scientific Imaging Problems}
%
%
%

\author[forthics]{Athanasia Panousopoulou\corref{mycorrespondingauthor}}
\cortext[mycorrespondingauthor]{Corresponding author}
\ead{apanouso@ics.forth.gr}

\author[cea]{Samuel Farrens}

\author[forthics,uoc]{Konstantina Fotiadou}

\author[safran]{Arnaud Woiselle}

\author[forthics,uoc]{Grigorios Tsagkatakis}

\author[cea]{Jean-Luc Starck}

\author[forthics,uoc]{Panagiotis Tsakalides}

\address[forthics]{Institute of Computer Science, Foundation of Research and Technology - Hellas, Greece}
\address[cea]{Laboratoire AIM, UMR CEA-CNRS-Paris 7, Irfu, Service d'Astrophysique, CEA Saclay, France}
\address[uoc]{Department of Computer Science, University of Crete, Greece}
\address[safran]{Safran Electronics \& Defense, France}


\begin{abstract}
Current trends in scientific imaging are challenged by the emerging need of integrating sophisticated machine learning with Big Data analytics platforms. This work proposes an in-memory distributed learning architecture for enabling sophisticated learning and optimization techniques on scientific imaging problems, which are characterized by the combination of variant information from different origins. We apply the resulting, Spark-compliant, architecture on two emerging use cases from the scientific imaging domain, namely: (a) the space variant deconvolution of galaxy imaging surveys (astrophysics), (b) the super-resolution based on coupled dictionary training (remote sensing). We conduct evaluation studies considering relevant datasets, and the results report at least 60\% improvement in time response against the conventional computing solutions. Ultimately, the offered discussion provides useful practical insights on the impact of key Spark tuning parameters on the speedup achieved, and the memory/disk footprint.\end{abstract}


\end{frontmatter}
%


\section{Introduction}
The last decade has been earmarked by the significant technological advances on both expensive and cost-effective instrumentation, which pervasively collects, processes, and communicates massive streams of information. The resulting deluge of manifold data has provided new pathways for ground-breaking scientific discoveries in various fields, ranging from neuroscience and system biology, to medicine and astrophysics, while challenging at the same time the computer engineering communities to enable the paradigm shift for accurate trends prediction over large-scale scientific datasets~\cite{Chen2016, marx2013}. 

The necessity of empowering distributed learning and inference over the petascales of scientific data is considered a game changer for distributed computing~\cite{Chen2014, Hu:2014} and respective management platforms, originally and vastly employed for retail and social networking services. Considering specifically the scientific imaging domain, the respective large-scale datasets (e.g., the Sloan Digital Sky Survey~\cite{sdss} and the upcoming Euclid space mission by the European Space Agency~\cite{euclid}) involve a rather small community of users, while confronting at the same time the challenge of sufficiently manipulating, and analyzing a significantly larger amount of information than the one considered in the social or Interner-based media~\cite{Huijse2014, Guo2015}. Interestingly, such datases possess significant ``V'' properties of Big Data beyond their voluminous character; their production rate can readily reach the magnitudes of exabytes/day (velocity), while reflecting on different points of origin (variety), and often demanding robust algorithms for noisy, incomplete or inconsistent data (veracity). Ultimately, while \emph{scientific imaging big data} can reflect on complex relationships (e.g., \cite{Song2017}),  they are expensive to create, difficult to maintain, and laborious to infer contextual information.

Performing analytics over scientific imaging big data corresponds to a problem far more complex than upgrading hardware infrastructures to meet the latest technological trends in hardware acceleration, or exchanging conventional machine learning techniques with their sophisticated deep learning counterparts; The computational complexity will continue to increase along with the scale of the input problem. Instead, in parallel to the synthesis of new models for coping with the manifold characteristics of imaging signals, the challenge is how the current knowledge in optimization and machine learning research can be optimally exploited for building efficient imaging data processing in large-scale settings. Specifically, the role of computational approaches that adopt a ``black-box'' approach should be revisited from the perspective of current and emerging trends in big data analytics platforms, in order to optimally reuse sophisticated techniques that have been designed for small-scale problems over scientific imaging big data, and ultimately enable a rapid pace of innovation for scientific discovery.

Notably, big data technology has been employed for addressing key processing steps in voluminous imaging problems (e.g., SciSpark~\cite{Wilson2016}, Kira~\cite{Zhang2016}). Even so, the migration from an academic implementation to a scalable solution over a distributed cluster of computers remains challenging; the correct execution require careful control and mastery of low-level details of the distributed environment~\cite{Xing2016}. At the same time, although many common learning algorithms are supported for big data, there is still both the practical need and the research interest to expose less-widely or new machine learning algorithms to the scientific imaging domain~\cite{Zhou2017}. In this work we address this gap by designing and developing an in-memory distributed learning architecture for applying sophisticated learning and optimization techniques on scientific imaging datasets. Specifically, we extend the work presented in~\cite{Panousopoulou2017}, which introduced a scheme compliant to the Apache Spark computing framework~\cite{Zaharia2016} for solving the problem of removing distortion from noisy galaxy images towards three directions, namely: (a) we formulate the overall distributed learning framework for scientific imaging problems with dedicated emphasis on addressing both the volume and variety of the input data, (b) we explore how the resulting architecture can be utilized for the efficient parallelization of the learning problem at hand, considering two use cases from the scientific imaging domain associated with astrophysics and remote sensing, (c) we evaluate the proposed architecture using realistic datasets for each use case and provide useful technical insights on the impact of key experimental parameters on the speedup achieved, and the memory/disk footprint. The results highlight the benefits of the proposed architecture in terms of speedup and scalability for both use cases, while offering practical guidelines for enabling analytics over scientific imaging big data. While employing commodity hardware as the cornerstone of the distributed environment, we achieve more than 60\% improvement in convergence rate terms with respect to conventional computing solutions, for both application scenarios. 

The remainder of the paper is organized as follows: in Section~\ref{relatedwork}, the recent trends on distributed computing tools for ScI-BD are outlined. The proposed distributed learning architecture is described in Section~\ref{arch}, while its instantiation for each use case along with the accompanying the evaluation studies are provided in Section~\ref{sec:usecases}. Conclusions are drawn in Section~\ref{sec:conclusions}.

\section{Background and Related Work}\label{relatedwork}
\subsection{Big Data platforms for large-scale learning}
Large-scale learning problems are typically empowered by distributed computing architectures, with the objective to disseminate the computational burden to a set of distributed resources. The resulting clusters of networked computing resources subsequently yield the underlying infrastructure, on top of which massive streams of raw data must be processed and transformed to meaningful information. As such, processing models and accompanying programming abstractions are essential for implementing the application logic and facilitating the data analytics chain.

Three main categories of process models can be identified, namely: (a) generic (e.g., MapReduce~\cite{Dean2008}, Dryad~\cite{Chen2014}) suitable for batch processing, (b) graph (e.g., GraphLab~\cite{Low2012}, GraphX~\cite{Xin2014}) accordingly expressing the relationship between data and computing tasks, and (c) streaming (e.g., S4~\cite{Neumeyer2010}, Storm~\cite{Toshniwal2014}) treating data as events and applying actor programming models. These programming models are instantiated on dedicated platforms for big data analytics. Depending on whether intermediate data are stored on the disk or in the memory, these platforms can support off-line, non-iterative, or on-line and iterative applications respectively. Hadoop and its variations~\cite{Wang2013, Ahuja2018} is considered the mainstream MapReduce implementation and enables large-scale non-iterative computations (e.g., sorting) by the means of replicating data on disks of multiple hosts. Despite its popularity, the main drawback of Hadoop is its response time; all intermediate data are stored in the disk, thereby causing significant latency during processing~\cite{Chen2014}.

An alternative to Hadoop, which allows in-memory analytics~\cite{Zhang:2015} over commodity hardware is the Spark framework~\cite{Zaharia2016}. Spark extends the MapReduce model by the use of 
an elastic persistence model, which provides the flexibility to persist these data records, either in memory, on disk, or both in memory and on disk. As such, Spark favors applications that need to read a batch of data multiple times, such as iterative processes met in machine learning and optimization algorithms. 

Spark adopts a structured design philosophy, which allows, among others, its extension towards Spark MLlib, an open-source machine learning library~\cite{Meng2016, Assefi2017} suitable for both unstructured and semi-structured data. The key characteristic of Spark MLlib is that it can assemble into a single pipeline a sequence of algorithms, while supporting automatic parameter tuning for all algorithms within the defined pipeline. Despite implementing more than fifty-five common algorithms for model training, 
 MLlib lacks of the essential data models for processing multi-dimensional arrays, such as the ones typically employed for expressing scientific and imaging data. To address this limitation, SparkArray~\cite{Wang2016} offers an extension of the Apache Spark framework that provides both a multi-dimensional array data model as well as a set of accompanying array operations on top of the Spark computing framework. Along similar lines, Misra et al.~\cite{Misra2018} identified that the primary bottleneck on matrix inversions over Spark and MLLib are the necessary matrix multiplications and, thus, proposed SPIN, a Spark-compliant distributed scheme for fast and scalable matrix inversion. 

\subsection{The positioning of Apache Spark in the distributed learning arena}
Overall, the in-memory philosophy of Spark, along with the advances on offering a repository for machine learning techniques and useful matrix operations, have substantially empowered its positioning in the distributed learning arena. A key issue recently addressed by the computing community~\cite{Xing2015, Zhang2017, Caino-Lores2018} relates to its performance against dedicated distributed learning platforms for scientific and imaging large-scale data. For instance, Petuum~\cite{Xing2015}, is a representative counterpart of Spark, and a specialized distributed machine learning framework relying on the principles of parameter server and state synchronous parallelism~\cite{Ho2013} for enabling high-performance learning. Compared to Spark, Petuum takes advantage of the iterative-convergence properties of machine learning programs for improving both the convergence rate and per-iteration time for learning algorithms. This results to improved speed-up for large-scale machine learning models with many parameters, at the expense of a less convenient model for programming and code deployment than the one offered by Spark. Considering disk-based distributed platforms for array data analytics, ArrayBench~\cite{Zhang2017} is a structured benchmark environment, over which a detailed analysis on the performance of Apache Spark against SciDB~\cite{Stonebraker2013} is performed. The thorough analysis on different aspects of time response and scalability over voluminous workflows representing gene and biological networks highlight both the superiority of the persistence models of offered by Spark for data-intensive analytics over SciDB, as well as its the importance of suitably tuning the configuration parameters for achieving optimal performance with respect to the memory and disk usage. Ultimately, Caino-Lores et al. present in~\cite{Caino-Lores2018} the reproduction of an iterative scientific simulator for hydrological data, originally implemented based on the principles of message passing interface (MPI), into Apache Spark and study the performance against both private and public cloud infrastructures. The insights therein offered highlight both the benefits of Spark in terms of easing data parallelization and underlying task management against the MPI-based approach, while revealing at the same time the impact of the configuration parameters in the final performance with respect to memory and stability. 

In parallel, current bibliography trends highlight the adoption of Spark for both key processing steps in large-scale imaging problems~\cite{Wilson2016, Palamuttam2015,Zhang2016, Zhang2015, Peloton2018}, as well as for parallelizing dedicated machine learning and optimization algorithms~\cite{Maillo2017, Arias2017, Huang2017,Makkie2018}. Specifically, with regard to imaging data management over Spark, SciSpark~\cite{Wilson2016, Palamuttam2015} pre-processes structured scientific data in network Common Format (netCDF) and Hierarchical Data Format (HDF). The result is a distributed computing array structure suitable for supporting iterative scientific algorithms for multidimensional data, with applications on Earth Observation and climate data for weather event detection. Another representative framework is Kira~\cite{Zhang2016, Zhang2015}, which leverages on Apache Spark for speeding-up the source extraction process in astronomical imaging, while outperforming high performance computing approaches for near real-time data analysis over astronomical pipelines. Finally, Peloton et al. synthesize a native Spark connector to manage arbitrarily large astronomical datasets in Flexible Image Transport System (FITS) formatin~\cite{Peloton2018}. The analysis therein provided indicates the capability of the proposed connector to automatically handle computation distribution and data decoding, thereby allowing the users to focus on the data analysis. 

With regard to employing Spark for the parallelization of dedicated machine learning and optimization algorithms, authors in~\cite{Maillo2017} and \cite{Arias2017} provide new solutions for in-memory supervised learning, elaborating on exact k-nearest neighbors classification and Bayesian Network Classifiers~\cite{Pearl2014}, respectively. In both cases, the efficacy in terms of time to completion and scalability is highlighted over annotated datasets with pre-defined features. Shifting towards imaging optimization problems, the authors in \cite{Huang2017} study the distributed asteroid detection as a complete framework over Spark and cloud computing that considers both the pre-processing of raw data, as well as parallelization of dedicated learning and optimization algorithms for asteroids detection. The resulting system provides the ability of both incrementally updating the new data from continuously observations, as well as the visual means for inspecting the position-linkage for the discovered asteroids. Finally, by employing a similar philosophy, Makkie et al. discuss in~\cite{Makkie2018} how the sparse characteristics of a dictionary learning algorithm for functional network decomposition can be implemented over Spark, for satisfying the desired scalability and reproducibility requirements of neuroimaging big data analysis.

The discussion thus far highlights the potential of in-memory computing frameworks, in general, and Spark, in particular, to enable the paradigm shift for large-scale analytics over scientific imaging. Even so, in its vast majority the related bibliography neither explicitly addresses how the multivariate characteristics of imaging datasets can be profiled in the distributed frameworks, nor extensively reports the role of Spark tuning parameters in the performance of distributed architectures. With respect to the current state of art, our contributions are the following:
 
\begin{itemize}
\item We elaborate on the system perspective for empowering distributed learning over large-scale and multivariate imaging datasets, and provide the respective distributed architecture based on Spark; 
\item We explore how the proposed architecture can be instantiated for two emerging imaging problems and respective state-of-art theoretical solutions, namely (a) the space variant deconvolution of noisy galaxy images (astrophysics), and (b) the super-resolution using dictionary learning (remote sensing);
\item We evaluate the resulting implementations using commodity hardware and realistic datasets, highlighting the relationship between the size of the input problem and computational capacity of the distributed infrastructure; 
\item We study the role of Spark tuning parameters (i.e., number of partitions and persistence model) on the memory and time performance, with detailed discussion in terms of speedup, scalability, memory/disk usage and convergence behavior;
\item Finally, we offer practical guidelines on the tuning of the distributed learning architecture for imaging problems, as derived from the evaluation procedure.
\end{itemize}

\section{Distributed Learning for Scientific Imaging Problems}\label{arch}

\subsection{Spark preliminaries}
\begin{figure*}[h!t]
\begin{tabular}{cc}
\small{(a)} & \small{(b)}\\
\includegraphics[scale=0.12]{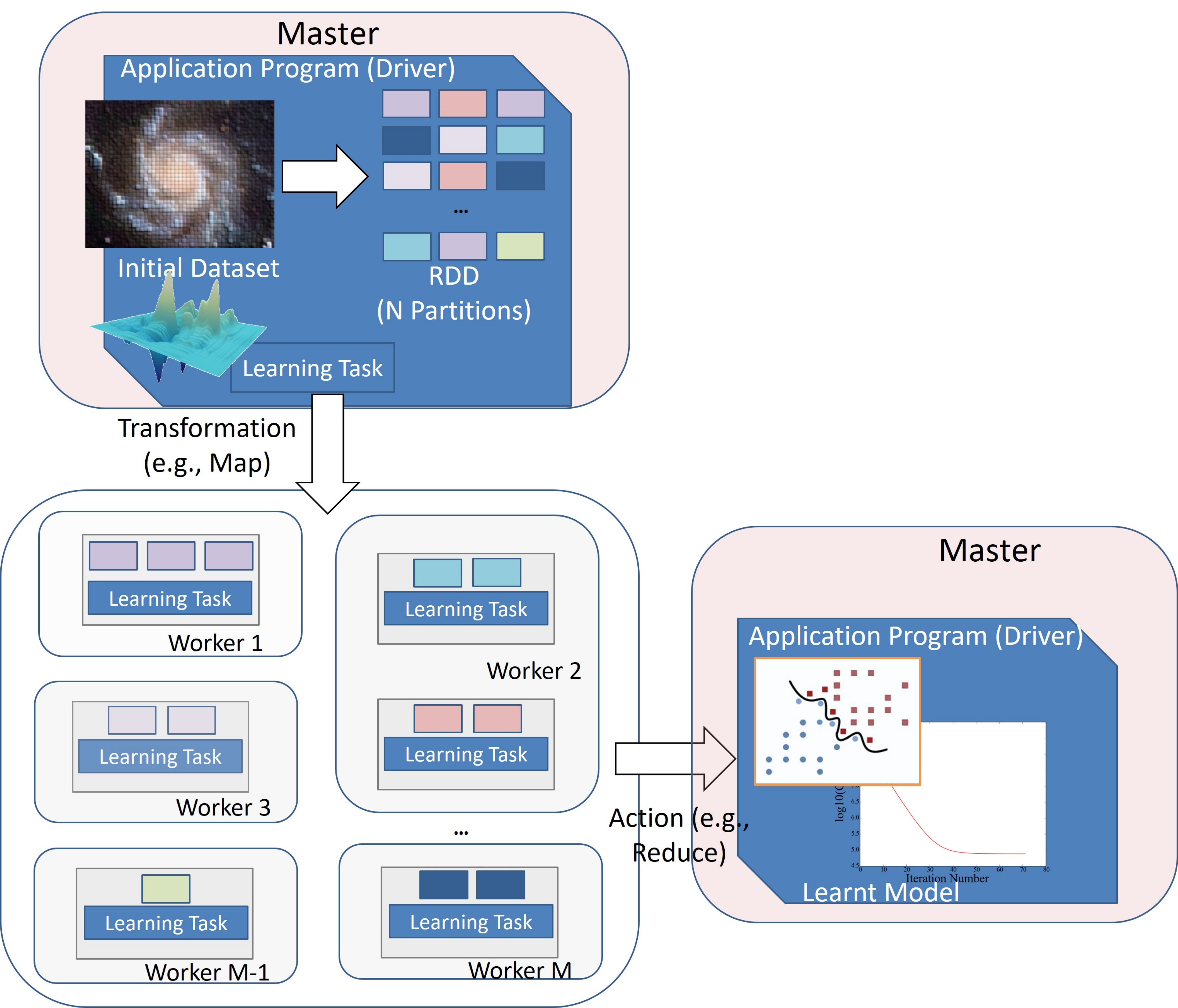} & \includegraphics[scale=0.3]{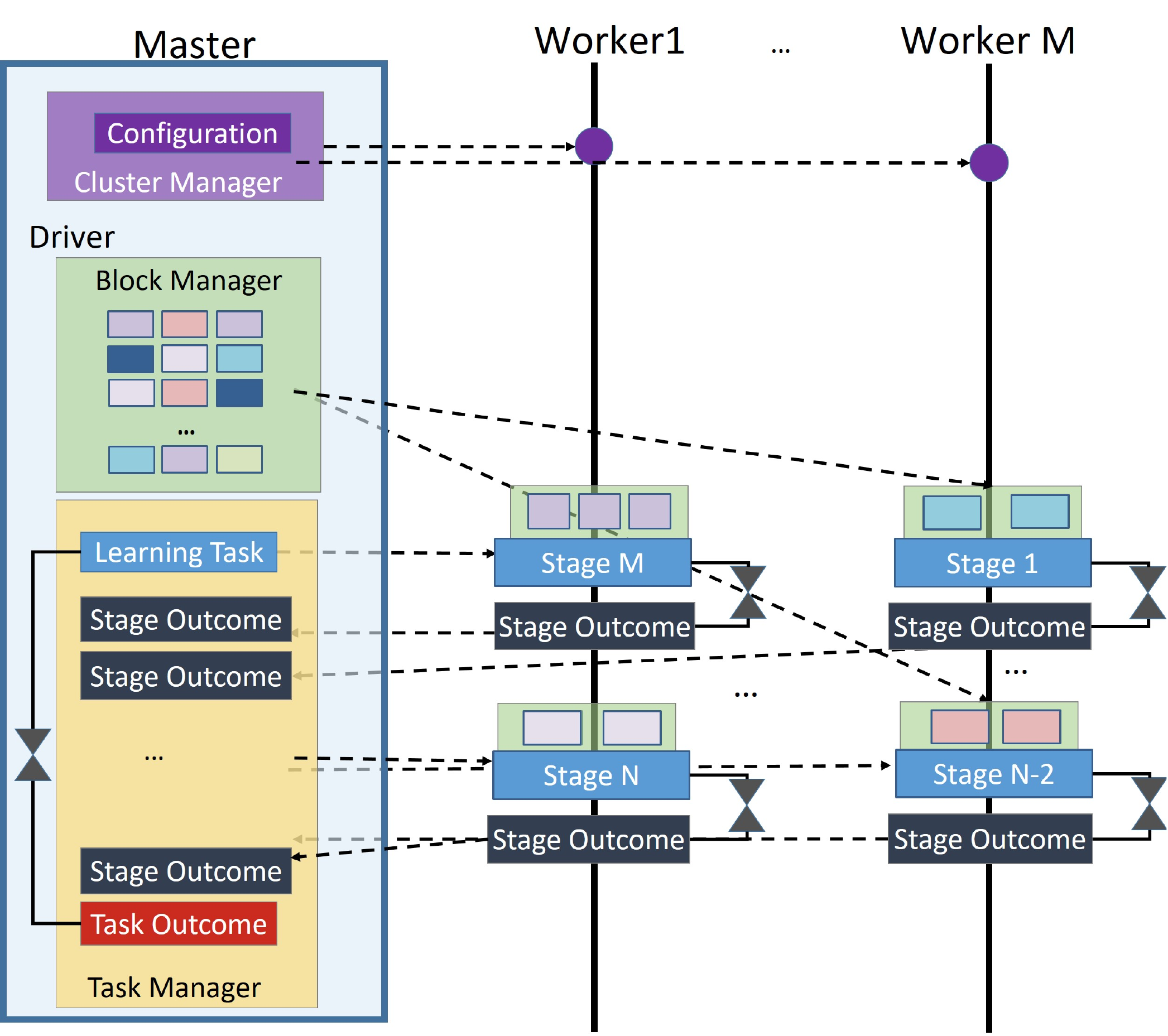}
\end{tabular}
\caption{Overview of the Spark: (a) the master-slave architecture for deploying and executing a distributed learning task, (b) the interaction between the master and the workers during the execution of a learning task.}\label{fig:spark_prem}
\end{figure*}

Spark organizes the underlying infrastructure into a hierarchical cluster of computing elements, comprised of a master and a set of $M$ workers (Fig.~\ref{fig:spark_prem}(a)). The master is responsible for implementing and launching the centralized services associated to the configuration and operation of the cluster, while the workers undertake the execution of the computing tasks across a large-scale dataset. The application program, which define the learning problem and the respective datasets, are submitted to the cluster through the \emph{driver} program, which is essentially the central coordinator of the cluster for executing the specific computations. The driver is responsible for partitioning the dataset into smaller partitions which are disseminated to the workers, and sequentially sending tasks to the workers. The workers perform the learning task on their assigned data chunks and report back to the driver with the status of the task and the result of the computation. 

The partitioning of the dataset relies on the Resilient Distributed Datasets (RDD)~\cite{Zaharia:2012}, which are defined as read-only collections of $N$ data records that is created through deterministic operations (e.g., map, group with respect to a key), on either the initial dataset, or another RDD. The resulting data blocks are parceled into the $M$ workers of the cluster. Each RDD is characterized by its lineage, which essentially conveys in a directed acyclic graph (DAG) enough information on how it was constructed from other RDD. As such, the lineage of an RDD can be utilized for reconstructing missing or lost partitions, without having to checkpoint any data. Through the driver, the application program can control how the initial data will be parallelized into one or multiple RDD, apply transformations on existing RDD. These transformations are lazy; RDD are only computed when they are used in \emph{actions}, which are operations that return a value to the application program or export data to a storage system. As such, a set of pipelined transformations of an RDD will not be executed until an action is commanded.

The execution of an application program comprised of a set of sequential computing tasks is performed in three distinct phases (Fig.~\ref{fig:spark_prem}(b)) namely: (a) the configuration of the operational parameters; (b) the parallelization of the datasets and subsequent records into RDD; (c) the assignment and execution of the learning tasks. During the second phase, the Block Manager at the driver caters the workers with missing data blocks needed for computations throughout the lifetime of the application program. The third phases fires with the request of performing a learning task in the form of an action on the RDD. The Task Manager service calculates the DAG of the RDD lineage and accordingly assigns the execution of the learning task to the workers, in the form $N$ stages. 
The result of the stage returns back to the driver program, and the Task Manager assigns another stage of the same learning task, until all stages have been completed. This procedure is repeated for the remaining learning tasks. Notably, if the successive learning tasks are not independent from each other, e.g., part of an iterative optimization process, the lineage of the RDD will gradually increase, with a direct impact on the memory requirements.

\subsection{Proposed architecture}
The flexibility of Spark is partially grounded on the adopted data-parallel model, which is implemented as a pipeline of RDD through the appropriate combination of transformation and actions. Even so, a single transformation / action can handle neither multiple RDD at the same time, nor RDD nested within each other. This characteristic limits the applicability of the Spark framework on learning schemes over imaging data, which commonly rely on the combination of variant information from different origins for removing noisy artifacts and extracting essential information. Considering such problems, one can think of heterogeneous imagery that correspond to the same spatial information (e.g., patches of noisy and reference images) to be jointly processed for solving single/multi-objective optimization problems. As such, a substantial volume of bundled imaging data should become readily available in iterative processes for enabling large-scale imaging analytics. 

\begin{figure*}[htbp]
\setlength{\abovecaptionskip}{0.5pt} 
\begin{center}
\begin{tabular}{cc}
\small{(a)} & \small{(b)}\\
\includegraphics[scale=0.09]{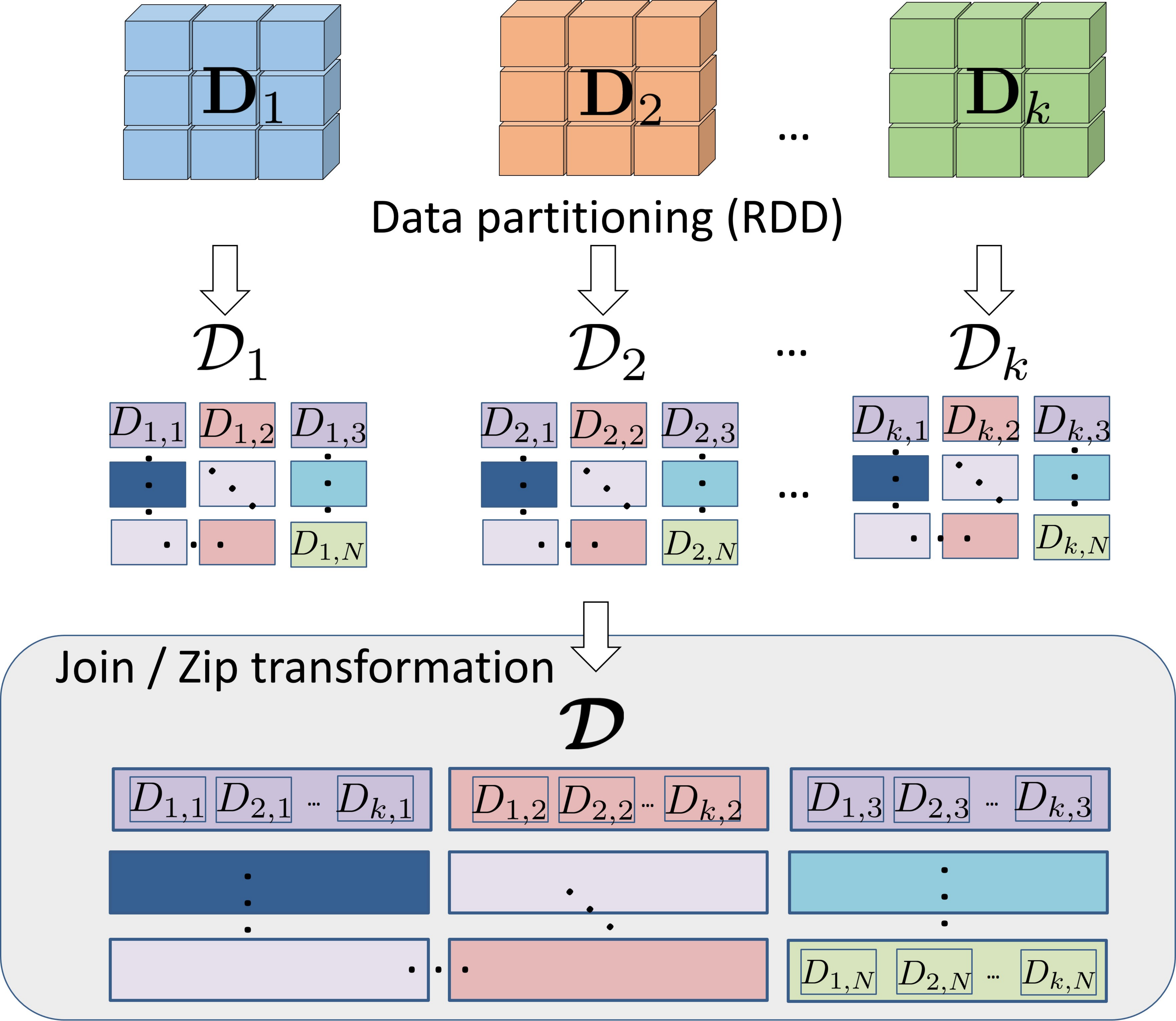} & \includegraphics[scale=0.09]{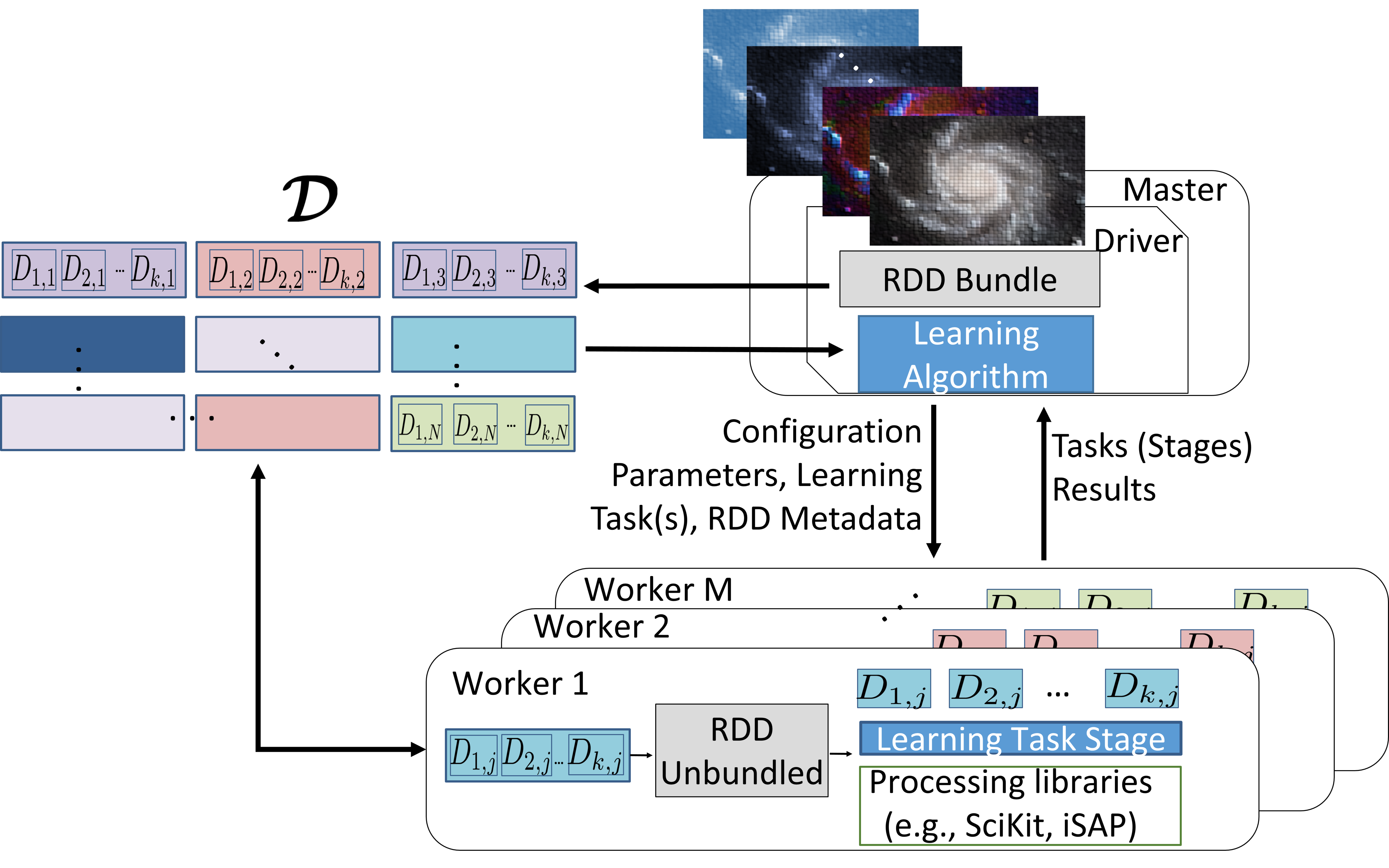}
\end{tabular}
\end{center}
\caption{(a) The creation of the bundled RDD $\boldsymbol{\mathcal{D}}$ over Spark, (b) the architecture of the proposed scheme.}\label{fig:arch1}
\end{figure*}

The herein proposed architecture considers the native RDD abstraction offered by Spark for performing distributed learning over bundled imaging datasets. Without loss of generality we consider that a set of $k$ imaging datasets and auxiliary structures (e.g., optimization variables, reference images), are modeled as multidimensional arrays $\mathbf{D}_{1}, \mathbf{D}_{2}, \ldots, \mathbf{D}_{k}$. For each $\mathbf{D}_{i}$, $i \in \{1,2,\ldots,k\}$ the driver program creates the respective RDD $\mathcal{D}_{i}$, by essentially defining the partitioning of $\mathbf{D}_{i}$ into $N$ data blocks $D_{i,j}$, where $j \in \{1,2, \ldots N\}$, i.e., $\mathcal{D}_{i} = \left[D_{i,1}; D_{i,2}; \ldots, D_{i,N}\right]$. Applying a join- or zip-type of transformation across all individual $\mathcal{D}_{i}$ results to the creation of the bundled RDD  $\boldsymbol{\mathcal{D}}$, which combines the parallelized set of the $k$ input datasets (Fig.~\ref{fig:arch1}(a)):
\[
\boldsymbol{\mathcal{D}}  =  \left[\mathcal{D}_{1}, \mathcal{D}_{2}, \ldots, \mathcal{D}_{k}\right]  =  \begin{bmatrix}
D_{1,1} & D_{2,1} &\dots & D_{k,1}\\
D_{1,2} & D_{2,2} & \dots & D_{k,2}\\
 \vdots & \vdots & \ddots & \vdots \\
D_{1,N} & D_{2,N} & \dots & D_{k,N}\\
\end{bmatrix}
\]

For the implementation of the proposed scheme over Spark both the learning problem and the $k$ respective datasets, are submitted to the master of the cluster through the driver program. The initial imagery datasets can be either locally available on the master or stored in a Spark-compliant distributed file system (e.g., HDFS). The combination of the initial datasets and their subsequent parallelization into $\boldsymbol{\mathcal{D}}$ are undertaken by the RDD Bundle Component located at the side of the driver (Fig.~\ref{fig:arch1}(b)). Any transformation defining parts of the learning algorithm that should consider the combination of the $k$ imaging datasets can in turn be utilized with the resulting bundled $\boldsymbol{\mathcal{D}}$ at the driver. When the respective action is fired, both the learning task and $\boldsymbol{\mathcal{D}}$ are parceled into the $M$ workers of the cluster. Prior the execution of each learning task on the workers, each partition $j$ of $\boldsymbol{\mathcal{D}}$ is separated into the original inputs $D_{1, j}, D_{2,j}, \ldots, D_{k,j}$ for performing the task at hand. This is performed by the RDD Unbundle Component at the side of each worker. The calculation of the result of each task from the cluster is consistent to the architecture presented in Fig.~\ref{fig:spark_prem}(b); each learning task is split into $N$ sequential computing stages, which are performed on different sets of $k$ data blocks by different workers. The result of each stage returns back to the driver program, which assigns another stage of the same learning task, until all stages have been completed. In case the learning task is part of an iterative calculation, any updated part of the $i$-th partitioned blocks $D_{i, 1}, D_{i,2}, \ldots, D_{i,N}$ is bundled back into $\boldsymbol{\mathcal{D}}$ for the next iteration. Once all learning tasks are completed, the result is either returned to the driver program or stored in the distributed file system, using the build-in actions of Spark. 

The proposed scheme essentially provides a Spark-native approach for applying pipelined RDD transformations over a set of different imagining datasets that should be jointly processed. Much like any other Spark-based architecture, the driver program defines the application-specific tasks, along with the input datasets for parallelization and bundling, and their partitioning scheme, by defining for instance the value of $N$.  In addition to the application-defined tasks, dedicated processing libraries (e.g., Astropy~\cite{Robitaille:2013}, iSAP~\cite{Fourt:2013}, SciKit-Learn~\cite{Pedregosa2011}) can be deployed at each worker for solving the learning problem at hand. Notably, the use of the RDD Bundle and RDD Unbundle components helps retaining the core principles of the original learning algorithm intact, thereby facilitating the re-usability of existing and novel approaches in machine learning, originally designed and evaluated for small-scale scenarios. Even so, as detailed in Section~\ref{sec:usecases}, the problem and data characteristics have a key role in instantiating the proposed scheme for designing and developing in-memory distributed sophisticated learning techniques for imaging big data.

\section{Use Cases}\label{sec:usecases}
The herein proposed distributed learning platform has been employed for addressing the large-scale challenges of two application scenarios in imaging and respective datasets, namely: (a) the space variant deconvolution of galaxy survey images, and (b) the joint dictionary training for image super-resolution over low- and high-resolution data, entailing video streaming in either different bands (hyperspectral) or standardized RGB coloring. The respective software libraries and datasets are publicly available at~\cite{papercode}.

\subsection{Astrophysics: Space variant deconvolution of galaxy survey images}\label{psf_usecase}
Even in ideal conditions, astronomical images contain aberrations introduced by the Point Spread Function (PSF) of the telescope. Therefore, obtaining accurate and unbiased properties of extended sources, such as galaxies, from these images is predicated on the ability to remove or compensate for the effects of the PSF.

The PSF describes the response of an imaging system to point sources. The PSF of an astronimical instrument can either be modelled, assuming complete knowledge of the imaging system, or measured from distant stars in the field, which can be approximated as point sources. However, removing the PSF, a process referred to as deconvolution, is a non-trivial problem given the presence of random noise in the observed images that prevents the use of analytical solutions. This problem is even more complicated for upcoming space surveys, such as the Euclid mission, for which the PSF will aditionally vary across the field of view \cite{Laureijs2011}. 

Currently, very few efficient methods exist for dealing with a space variant PSF. 
 Nevertheless, an elegant solution is the concept of Object-Oriented Deconvolution~\cite{Starck2000}, which assumes that the objects of interest can first be detected using software like SExtractor~\cite{Bertin:96} and then each object can be independently deconvolved using the PSF associated to its center. This can be modeled as $\mathbf{Y} = \mathcal{H}(\mathbf{X}) + \mathbf{N}$, 
\noindent where $\mathbf{Y} =[\mathbf{y}^0, \mathbf{y}^1, \cdots, \mathbf{y}^n]$ is a stack of observed noisy galaxy images, $\mathbf{X} = [\mathbf{x}^0, \mathbf{x}^1, \cdots, \mathbf{x}^n]$ is a stack of the true galaxy images, $\mathbf{N} = [\mathbf{n}^0, \mathbf{n}^1, \cdots, \mathbf{n}^n]$ is the noise corresponding to each image and $\mathcal{H}(\mathbf{X}) = [\mathbf{H}^0\mathbf{x}^0, \mathbf{H}^1\mathbf{x}^1, \cdots, \mathbf{H}^n\mathbf{x}^n]$ is an operator that represents the convolution of each galaxy image with the corresponding PSF for its position.

In order to solve a problem of this type one typically attempts to minimize some convex function such as the least squares minimization problem:

\begin{equation}
   \begin{aligned}
        & \underset{\mathbf{X}}{\text{argmin}}
        & \frac{1}{2}\|\mathbf{Y}-\mathcal{H}(\mathbf{X})\|_2^2,
   \end{aligned}    
\label{eq:l2min}
\end{equation}

\noindent which aims to find the solution $\hat{\mathbf{X}}$ that gives the lowest possible residual ($\mathbf{Y} - \mathcal{H}(\hat{\mathbf{X}})$). This problem is ill-posed as even the tiniest amount of noise will have a large impact on the result of the operation. Therefore, to obtain a stable and unique solution to Eq.~(\ref{eq:l2min}), it is necessary to regularize the problem by adding additional prior knowledge of the true images. Farrens et al. proposed in \cite{farrens:17} two alternatives for addressing the regularization issues, namely: (a) a sparsity approximation, (b) a low-rank approximation. Briefly, the sparsity approximation imposes that the desired solution should be sparse when transformed by a given dictionary. In the case of galaxy images, this dictionary corresponds to an isotropic wavelet transformation \cite{starck:book15}, and Eq.~(\ref{eq:l2min}) becomes:

\begin{equation}
 \begin{aligned}
    C(\mathbf{X}) =  & \underset{\mathbf{X}}{\text{argmin}}
        & \frac{1}{2}\|\mathbf{Y}-\mathcal{H}(\mathbf{X})\|_2^2 + \|\mathbf{W}^{(k)}\odot\Phi(\mathbf{X})\|_1
        & & \text{s.t.}
        & & \mathbf{X} \ge 0
    \end{aligned}
\label{eq:psf_sparse}
\end{equation}

    %
\noindent where: $\|\bullet \|_{2}^{2}$ denotes the Frobenius norm; the $\Phi$ operator realizes the isotropic undecimated wavelet transform without the coarse scale; $\mathbf{W}^{(k)}$ is a weighting matrix related to the standard deviation of the noise in the input images; $\odot$ denotes the Hadamard (entry-wise) product; $k$ is a reweighting index, necessary to compensate for the bias introduced by using the $l$1-norm.

The low-rank approximation, on the other hand, simply assumes that with a sufficiently large sample of galaxy images some properties of these objects should be similar and therefore a matrix containing all of the image pixels will not be full rank \cite{farrens:17}. In this case, Eq.~(\ref{eq:l2min}) becomes: 

\begin{equation}
 \begin{aligned}
         C(\mathbf{X}) =  & \underset{\mathbf{X}}{\text{argmin}}
        & \frac{1}{2}\|\mathbf{Y}-\mathcal{H}(\mathbf{X})\|_2^2 + \lambda\|\mathbf{X}\|_*
        & & \text{s.t.}
        & & \mathbf{X} \ge 0
   \end{aligned}
\label{eq:psf_lowr}
\end{equation}

\noindent where $\lambda$ is a regularization control parameter and $\| \bullet \|_{*}$ denotes the nuclear norm.

As explained in~\cite{farrens:17}, both of these these techniques can be used to improve the output of the optimisation problem and hence the quality of the image deconvolution, which is considered a key factor for the success of the Euclid mission.

\subsubsection{Parallelization using the distributed learning architecture}
\begin{figure*}[htbp]
\begin{center}
\includegraphics[scale=0.2]{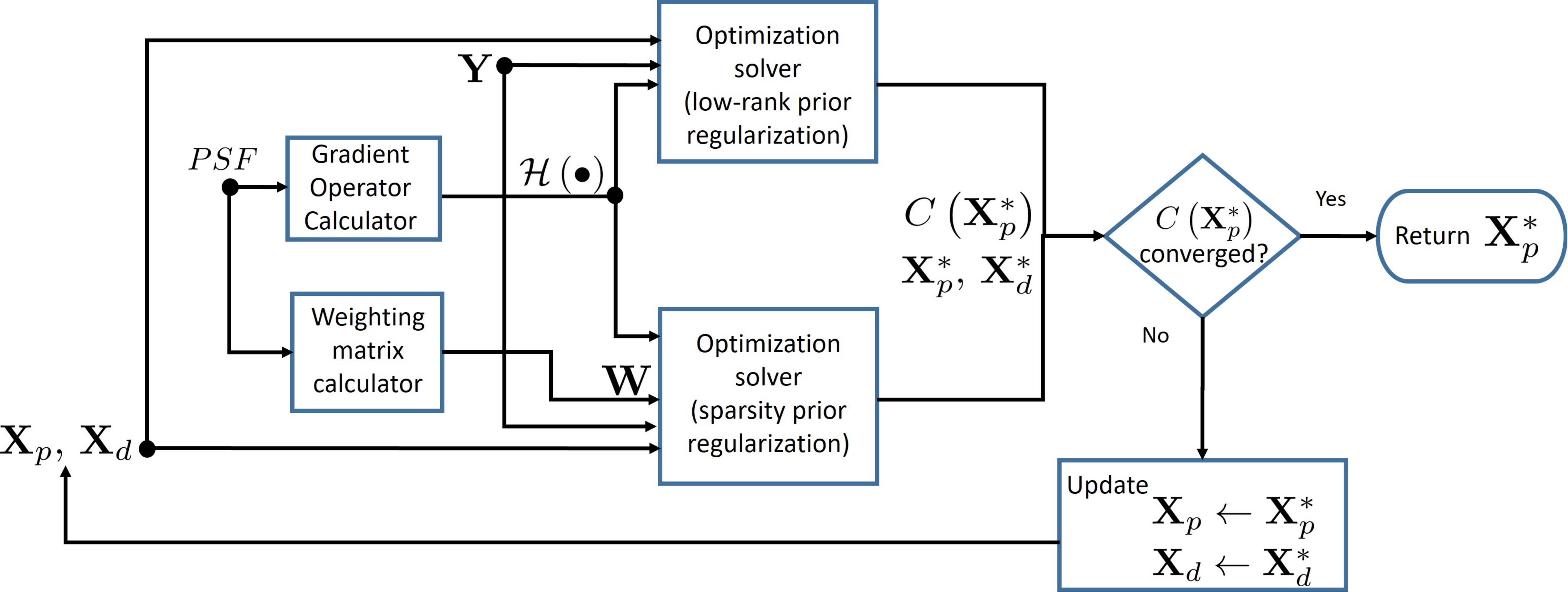}
\end{center}
\caption{The flow diagram for the sequential implementation of the space variant deconvolution of noisy galaxy images.}\label{figforth:psf_flow1}
\end{figure*}

A primal-dual splitting technique \cite{condat:13} can be adopted for solving both Eq.~(\ref{eq:psf_sparse}) and Eq.~(\ref{eq:psf_lowr}), taking into account the inherent sparse or low-rank properties, respectively. The sequential approach for implementing it reflects on an iterative optimization process. During each iteration step all requested input, i.e., noisy data ($\mathbf{Y}$), PSF data, primal ($\mathbf{X}_{p}$) and dual ($\mathbf{X}_{d}$) optimization variables, and weighting matrix ($\mathbf{W}$) -for the case of the sparsity-prior regularization- are jointly fed to the solver in order to calculate the value $C(\mathbf{X_{p}})$ of the cost function described by either Eq.~(\ref{eq:psf_sparse}) or  Eq.~(\ref{eq:psf_lowr}), as illustrated in Fig.~\ref{figforth:psf_flow1}.

This approach would be inefficient when the size of the noisy input data and respective space-variant PSF increases. In order to consider the necessary interaction between the different inputs for solving this optimization problem, we herein propose Algorithm~\ref{psf_parallel} for distributing the optimization phase according to the proposed learning architecture. 

\begin{algorithm*}[ht!]
\caption{The PSF algorithm parallelization (\colorbox{red!20}{steps} are performed on the driver, while \colorbox{green!10}{steps} are performed over the cluster).}
\label{psf_parallel}
\begin{footnotesize}
\KwData{The data $\mathbf{Y}$, the respective PSF, the maximum number of iterations $i_{\max}$, and the cost tolerance $\epsilon$. Typically: $i_{\max}=$300, $\epsilon=$10$^{-4}$}
\KwResult{The estimated images $\mathbf{X}_{p}^{*}$ that minimize Eq.~(\ref{eq:psf_sparse}) or Eq.~(\ref{eq:psf_lowr}).}
\nl \HiLiDR Initialize $\mathbf{X}_{p}$, $\mathbf{X}_{d}$ and extract the operator $\mathcal{H}(\bullet)$.\\ 
\nl \HiLiDR Define the RDD for $\mathbf{Y}$, PSF, $\mathbf{X}_{p}$, $\mathbf{X}_{d}$ as $\mathcal{D}_{\mathbf{Y}}$, $\mathcal{D}_{PSF}$, $\mathcal{D}_{\mathbf{X}_{p}}$, $\mathcal{D}_{\mathbf{X}_{d}}$, respectively, with $N$ partitions per RDD.\\
\nl \HiLiDR \uIf{Sparsity~regularization:}{ \HiLiCL Apply the weighting matrix calculator to $\mathcal{D}_{PSF}$: $\mathcal{D}_{\mathbf{W}} = \mathcal{D}_{PSF}.\text{\texttt{map}}(\text{\texttt{lambda} } x: \mathbf{W}^{(k)}(x))$.\\
\nl \HiLiCL Create the RDD bundle: $\boldsymbol{\mathcal{D}} = \mathcal{D}_{\mathbf{Y}}.\text{\texttt{zip}}(\mathcal{D}_{PSF}).\text{\texttt{zip}}(\mathcal{D}_{\mathbf{W}}).\text{\texttt{zip}}(\mathcal{D}_{\mathbf{X}_{p}}).\text{\texttt{zip}}(\mathcal{D}_{\mathbf{X}_{d}})$.\\}
\nl \HiLiDR \ElseIf{Low-rank regularization:}{ \HiLiCL Create the RDD bundle: $\boldsymbol{\mathcal{D}} = \mathcal{D}_{\mathbf{Y}}.\text{\texttt{zip}}(\mathcal{D}_{PSF}).\text{\texttt{zip}}(\mathcal{D}_{\mathbf{X}_{p}}).\text{\texttt{zip}}(\mathcal{D}_{\mathbf{X}_{d}})$.\\}
\nl \HiLiDR\For{$i:1:i_{\max}$}{ 
\nl \HiLiCL Update $\mathcal{D}_{\mathbf{X}_{p}}$, $\mathcal{D}_{\mathbf{X}_{d}}$ in $\boldsymbol{\mathcal{D}}$ using~\cite{condat:13}:  $\boldsymbol{\mathcal{D}}=\boldsymbol{\mathcal{D}}.\text{\texttt{map}}(\text{\texttt{lambda}}x: \text{Update } x)$.\\
\nl \HiLiCL Update $C(\mathbf{X}_{p}^{*})$ (Eq.~(\ref{eq:psf_sparse}) or Eq.~(\ref{eq:psf_lowr}), depending on sparsity or low-rank regularization):\\
\nl \HiLiCL $C(\mathbf{X}_{p}^{*}) = \boldsymbol{\mathcal{D}}.\text{\texttt{map}}(\text{\texttt{lambda} } x:$ $\text{Calculate }C(x)).\text{\texttt{reduce}(\texttt{lambda} }x,y: x+y)$.\\
\nl \HiLiDR\If{ $C(\mathbf{X}_{p}^{*}) \le \epsilon$}{
\nl \HiLiDR break \\
}
}
\nl \HiLiCL Save $\boldsymbol{D}$ to disk or distributed file system and return $\mathbf{X}_{p}^{*}$ to driver.\\
\end{footnotesize}
\end{algorithm*}
\setlength{\textfloatsep}{0pt}

Algorithm~\ref{psf_parallel} entails the parallelization of $\mathbf{Y}$, PSF data, $\mathbf{X}_{p}$, $\mathbf{X}_{d}$ into $\mathcal{D}_{\mathbf{Y}}$, $\mathcal{D}_{PSF}$, $\mathcal{D}_{\mathbf{X}_{p}}$, $\mathcal{D}_{\mathbf{X}_{d}}$ respectively on the side of the driver program. When the sparsity-prior regularization solution is adopted, the inherited dependency of the weighting matrix $\mathbf{W}$ on the PSF data, leads to the transformation of $\mathcal{D}_{PSF}$ to the corresponding weighting data blocks $\mathcal{D}_{\mathbf{W}}$. All requested input is in turn compressed into $\boldsymbol{\mathcal{D}}$, which essentially contains tuples of the form $<\mathbf{Y}, PSF, \mathbf{W}, \mathbf{X}_{p}, \mathbf{X}_{d}>$ (i.e., $k=$5, sparsity solution) or $<\mathbf{Y}, PSF, \mathbf{X}_{p}, \mathbf{X}_{d}>$ (i.e., $k=$4, low-rank solution). The resulting RDD $\boldsymbol{\mathcal{D}}$ is used to calculate the updated value of the optimization variable $\mathbf{X}^{*}_{p}$ based on the sparsity prior (Eq.~(\ref{eq:psf_sparse})) or low-rank (Eq.~(\ref{eq:psf_lowr})) regulization on each worker. The firing of a \emph{reduce} action, triggers the value of the cost function $C(\mathbf{X}^{*}_{p})$. This process that relies on the the interaction between the driver and the workers (map to $\boldsymbol{D}$ $\rightarrow$ reduce to $C(\mathbf{X}^{*}_{p})$) is repeated until either the value of $C(\mathbf{X}^{*}_{p})$ converges to $\epsilon$, or until the maximum number of iterations is reached. The resulting stack of stack of the true galaxy images $\mathbf{X}^{*}_{p}$ is directly saved on the disk of the driver program.

\subsubsection{Evaluation Studies}\label{sec:psf_studies}
The evaluation studies emphasize on the performance of the PSF algorithm parallelization implemented over the herein proposed learning architecture. The data used for this work consists of two samples of 10,000 and 20,000 simulated $41\times 41$ pixel postage stamps, respectively. Each stamp contains a single galaxy image obtained from the Great3 challenge \cite{mandelbaum:14}. These images were convolved with a Euclid-like spatially varying and anisotropic PSF and various levels of Gaussian noise were added to produce a simplified approximation of Euclid observations. In total there are 600 unique PSFs~\cite{kuntzer:16}, which are down-sampled by a factor of 6 to avoid aliasing issues when convolving with the galaxy images~\cite{cropper:13}.

The distributed learning architecture features Spark 2.1.0~\cite{apache2.1.0:2017}, deployed over a cluster of $M=$5 workers. The driver allocates 8GB RAM, while 4 out of 5 workers allocate 2.8GB RAM and 4 CPU cores. The fifth worker allocates 2.8GB RAM and 8 CPU cores, thereby yielding in total 24 CPU cores and 14GB RAM. With regard to the RDD persistence model, for all experiments conducted we considered the default storage level (memory-only), and as such, RDDs are persistent in memory. In case this exceeds the memory availability, some partitions will not be cached and will be recomputed on the fly each time they are needed. 

The evaluation procedure emphasizes on the time performance with respect to the sequential implementation\footnote{\href{https://github.com/sfarrens/psf}{\textcolor{blue}{https://github.com/sfarrens/psf}}} in terms of speedup, execution time per optimization loop, and scalability; the memory usage; the convergence behavior of the cost function. The key experimental parameters are the solution approach (sparsity or low-rank), and the data size (10,000 or 20,000 stack of galaxy images), with respect to the number of partitions $N$ partitions per RDD. We herein consider $N=\{2x, 3x, 4x, 6x\}$, where $x$ corresponds to the total number of cores available for parallel calculations, i.e. $x=$24. Notably, considering other parallel computing frameworks for comparison (e.g., Hadoop) is beyond the scope of this work, as they are either unsuitable for the essential iterative computations typically met in learning imaging problems, or focus on the extraction of astronomical imaging data.

\begin{figure*}[ht!]
\begin{center}
\begin{tabular}{cc}
\small{(a)} & \small{(b)}\\
\includegraphics[scale=0.25]{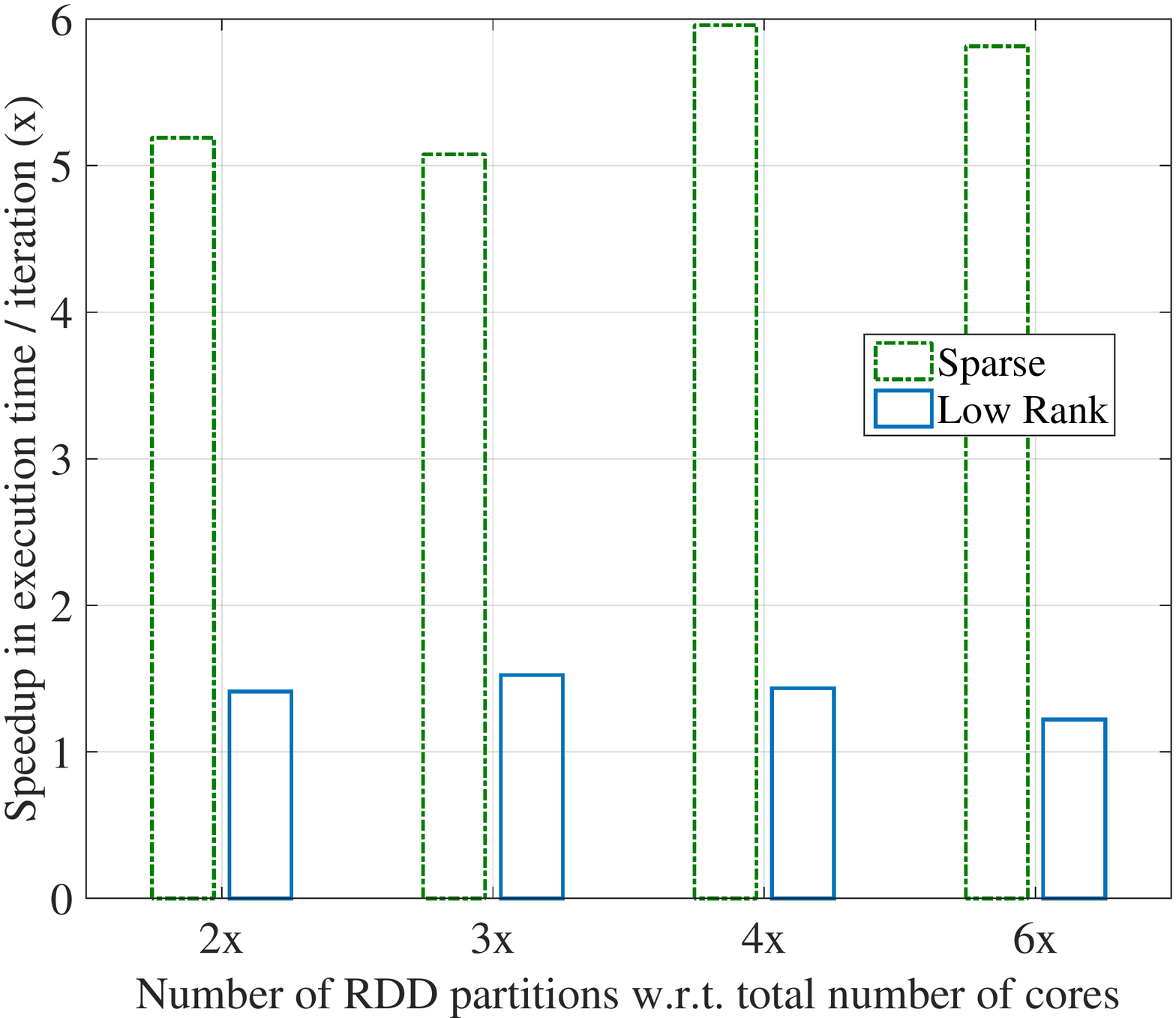} & \includegraphics[scale=0.25]{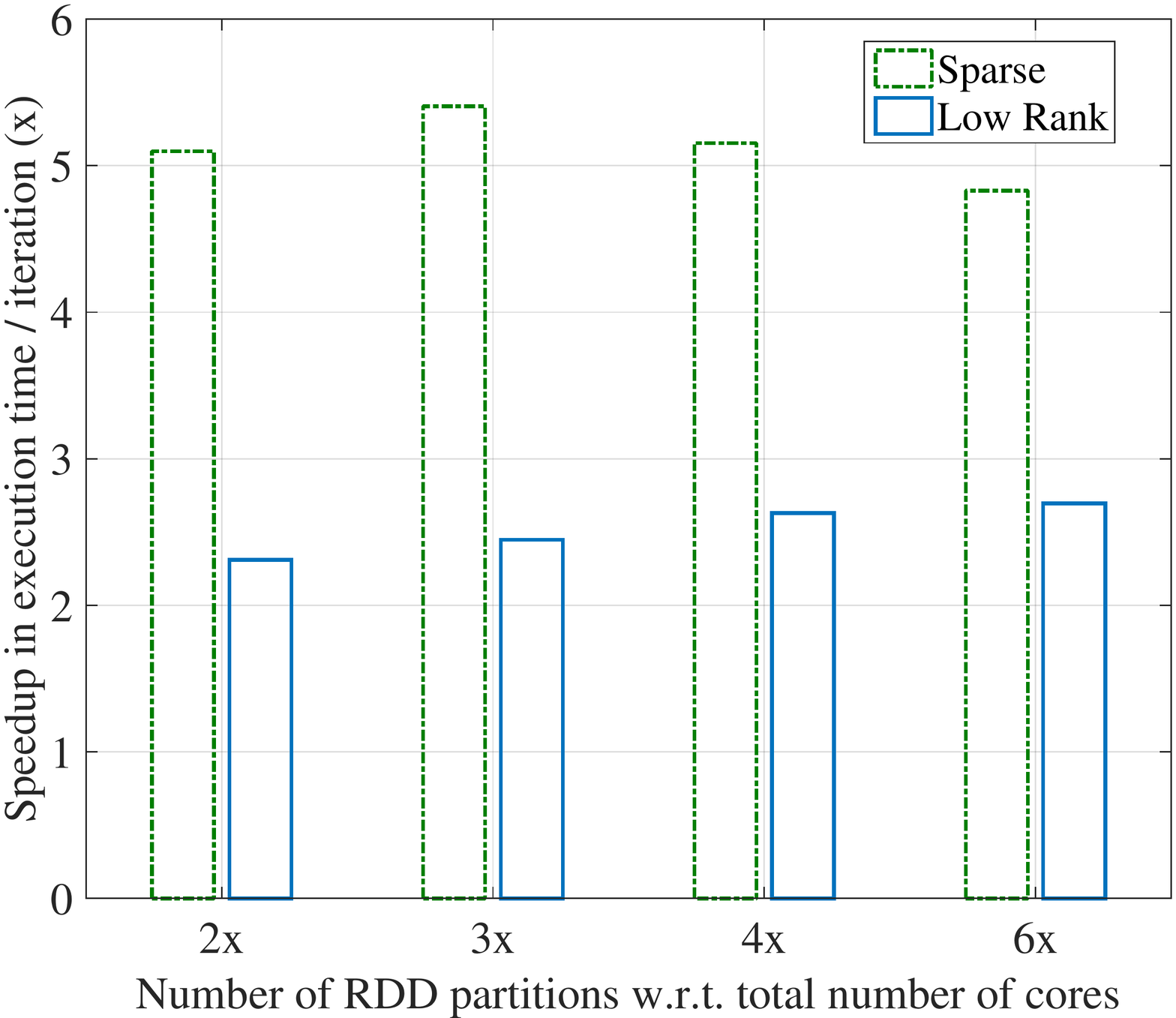}\\
\small{(c)} & \small{(d)}\\
\includegraphics[scale=0.25]{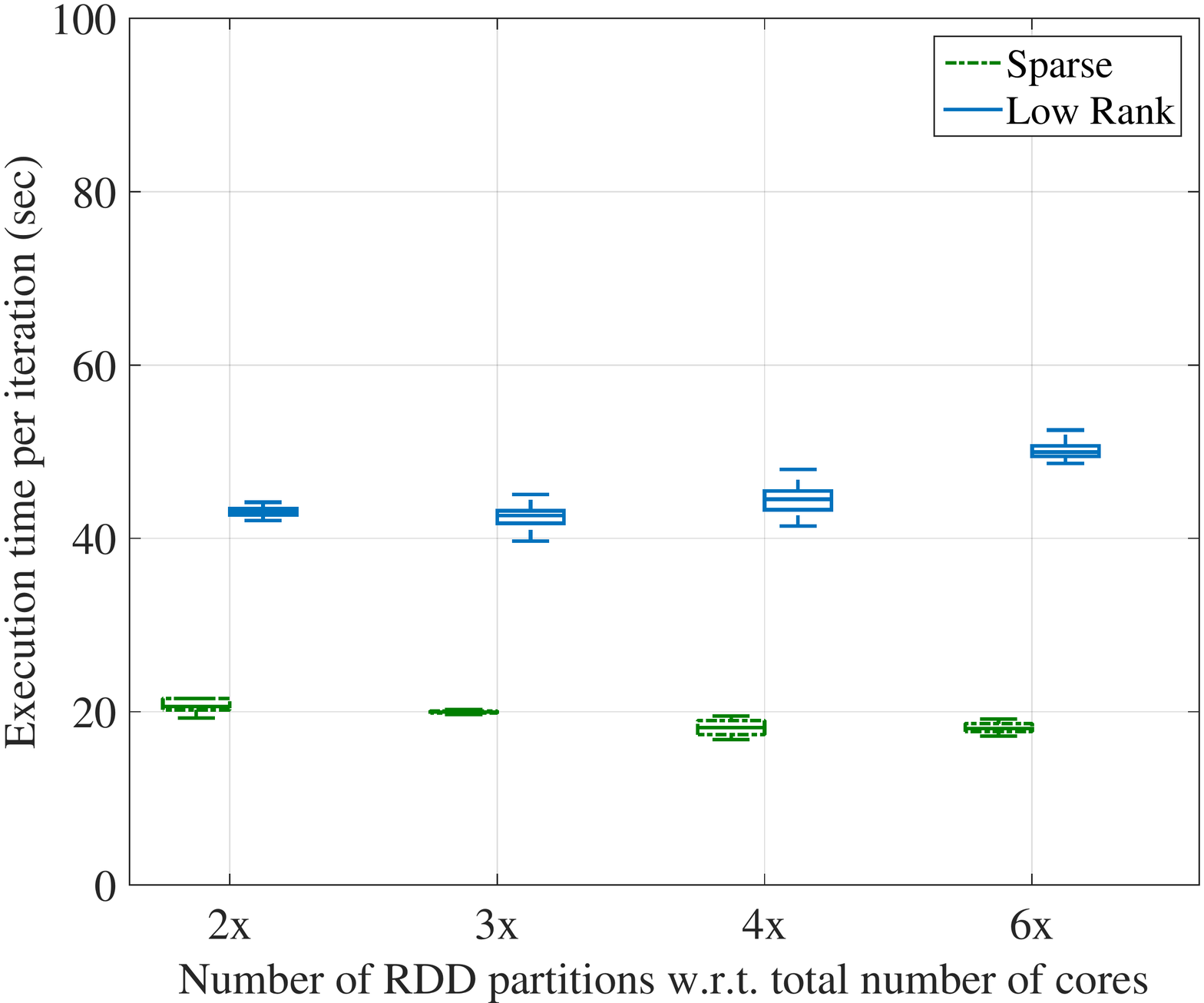} & \includegraphics[scale=0.25]{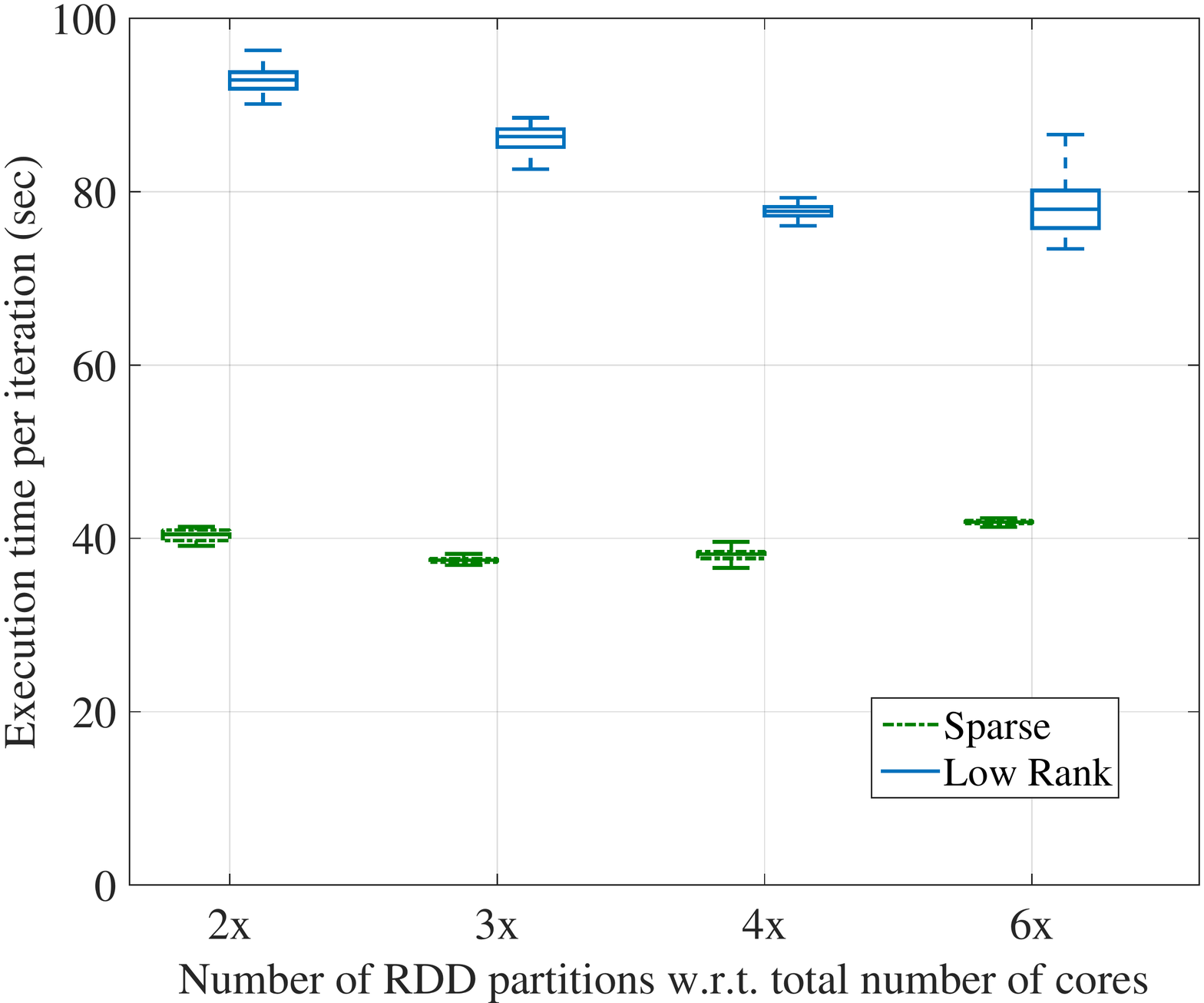}
\end{tabular}
\end{center}
\caption{Time performance for the parallelization of the PSF deconvolution based on the proposed architecture: (a) speedup for the stack of 10,000 images, (b) speedup for the stack of 20,000 images, (c) time execution per optimization loop for the stack of 10,000 images, (d) time execution per optimization loop for the stack of 20,000 images.}\label{fig:psf_speedup}
\vspace{0.5cm}
\end{figure*}
\noindent \textbf{Time performance.} Figure~\ref{fig:psf_speedup} presents the time performance achieved for both sparse- and low rank-based solutions in terms of speedup (Fig.~\ref{fig:psf_speedup}(a) for the stack of 10,000 images, and Fig.~\ref{fig:psf_speedup}(b) for the stack of 20,000 images), and execution time per iteration of the optimization process for the calculation of $C(\mathbf{X}_{p}^{*})$ (Fig.~\ref{fig:psf_speedup}(c) for the stack of 10,000 images, and Fig.~\ref{fig:psf_speedup}(d) for the stack of 20,000 images). The initial observation to make is that the herein proposed technique offers increased speed up. Greater improvement is achieved for the case of the sparsity-based solution (speedup $\ge$ 5$x$) than for the low-rank based solution (speedup $\in [1.2, 2.5]$) for all cases of $N$ examined and for both sizes of datasets considered (Fig.~\ref{fig:psf_speedup}(a) and (b)). This is due to the nature of the problem and the adopted solution; the sparsity-based solution has inherent parallelizable properties since the algorithm can work independently on different portions of the dataset. By contrast, the low-rank based solution performs SVD over the entire stack of images, and as such the parallelized data need to be reassembled at the driver program for the calculation of $\mathbf{X}_{p}^{*}$-$\mathbf{X}_{d}^{*}$. Nevertheless, as the number of noisy images increases from 10,000 to 20,000, the speedup of the low-rank solution improves to more than 2$x$ for all cases of $N$ considered. This is consistent to nature of the Spark framework, and highlights that the scale of an input imaging dataset is relevant to the capacity of the cluster, and the demands of the solving approach; the overhead introduced by Spark for the distributed calculations may hinder the overall performance, when the dataset is relatively small compared to the computational capacity of the cluster.
 
With respect to the time execution per optimization loop (Fig.~\ref{fig:psf_speedup}(c) and Fig.~\ref{fig:psf_speedup}(d)) we observe that the distributed architecture yields a stable performance, since for all combinations of solution approach, data size, and level of $N$, the time execution has limited statistical dispersion between the 25-th and 75-th percentiles. The impact of the parameter $N$ on the time performance is more evident for the low-rank solution, exhibiting a contradictory pattern; for the case of 10,000 images (Fig.~\ref{fig:psf_speedup}(c)) as the number $N$ of partitions per RDD increases (2$x$ $\rightarrow$ 6$x$), the median execution time per iteration is increased by approximately 8 secs ($\sim$42sec $\rightarrow$ 50sec). This is consistent to the speedup results, and suggest that the increase of $N$ implies more data chunks of smaller size need to be exchanged among the workers, thereby introducing unnecessary shuffling overhead due to the SVD computations. By contrast, when 20,000 images are considered (Fig.~\ref{fig:psf_speedup}(c)) the increase of $N$ between 2$x$ $\rightarrow$ 6$x$ results to a drop of the median execution time per iteration by approximately 10 secs ($\sim$98sec $\rightarrow$ 88sec). In this case, the partitioning into more data chunks implies less burden on memory per task, which substantially compensates any shuffling overhead. 

\begin{figure}[ht!]
\setlength{\abovecaptionskip}{0.5pt} 
\begin{center}
\includegraphics[scale=0.25]{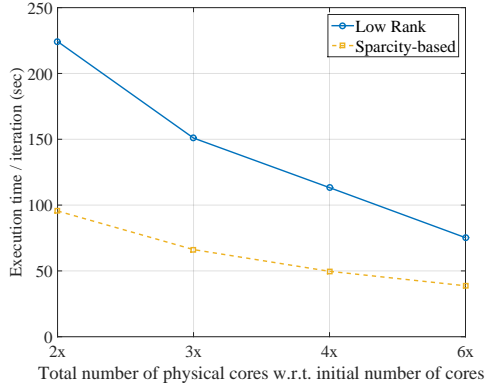}
\end{center}
\caption{Scalability of exeuction time per iteration for the stack of 20,000 images.}\label{fig:psf_scalability}
\end{figure}

Ultimately, scalability aspects on deploying Algorithm~\ref{psf_parallel} over the cluster infrastructure are illustrated in Fig~\ref{fig:psf_scalability} for the stack of 20,000 images and $N=$ 4$x$, for both sparse and low-rank approaches. As expected, as the total number of cores increases with respect to the cores considered for the evaluation of the sequential approach (i.e., 4 cores), the distributed learning approach offers substantial improvements in terms of time performance. Specifically, increasing the number of available cores in the cluster from 2$x$ to 6$x$ results into a 50\% and 65\% improvement for the sparse and the low-rank approach, respectively. 

\begin{figure*}[ht!]
\begin{center}
\begin{tabular}{cc}
\small{(a)} & \small{(b)}\\
\includegraphics[scale=0.25]{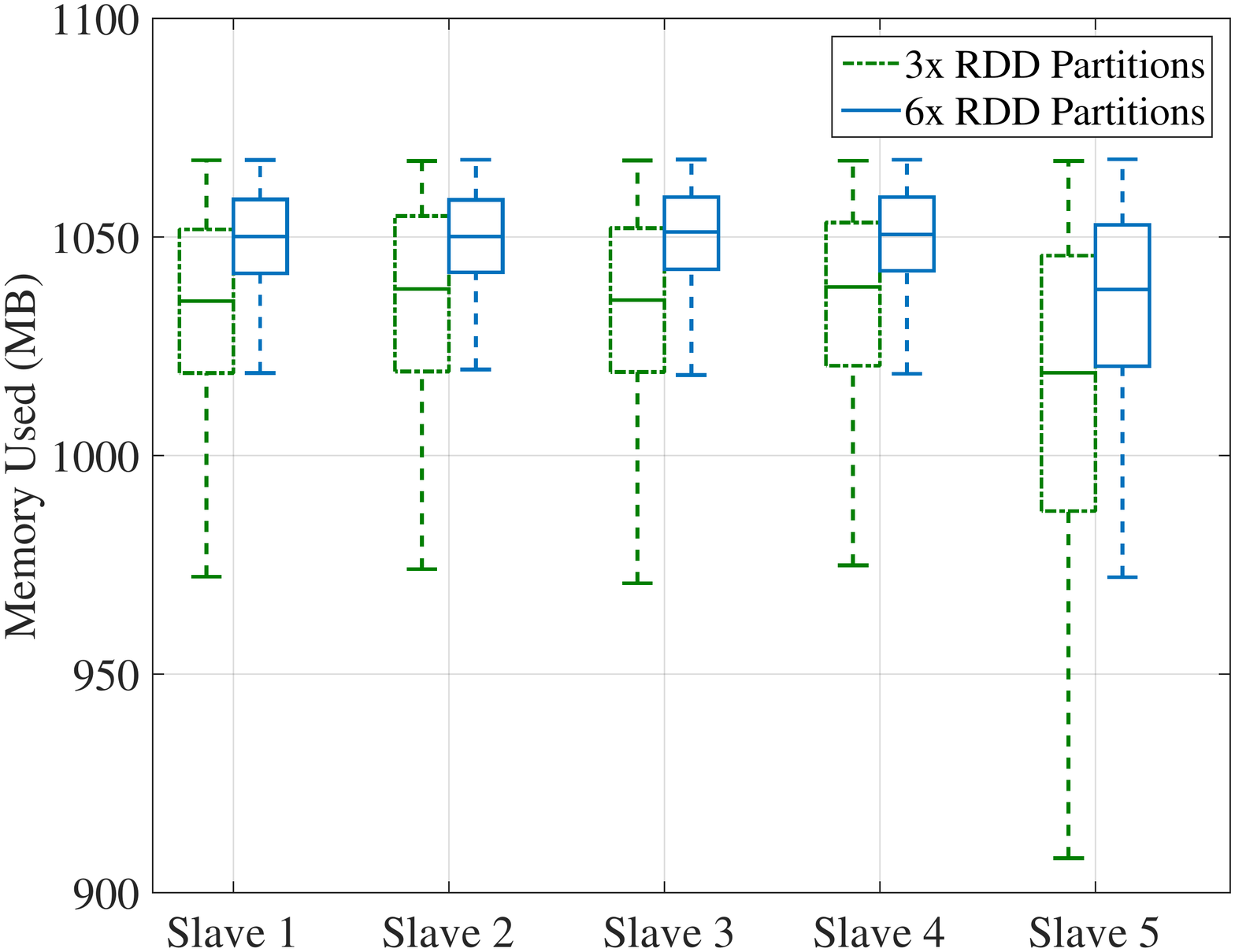} & \includegraphics[scale=0.25]{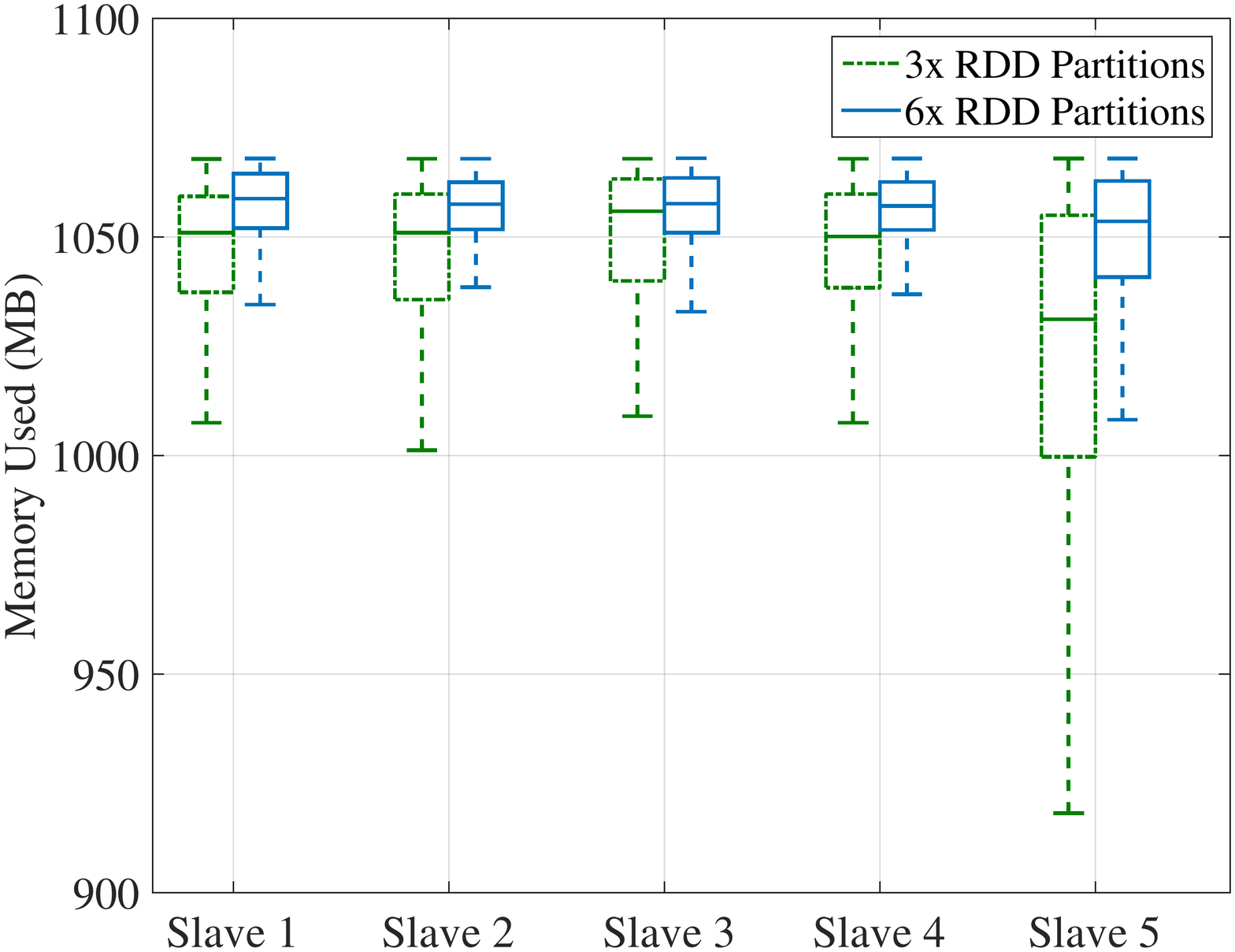}
\end{tabular}
\end{center}
\caption{Memory usage per slave when the stack of 20,000 images is considered for (a) sparsity based solution , (b) low-rank based solution.}\label{fig:psf_memory}
\vspace{0.5cm}
\end{figure*}

\noindent \textbf{Memory Usage.} Figure~\ref{fig:psf_memory} presents the memory usage per worker throughout the experiment duration considering the stack of 20,000 images and $N=\{$3$x$, 6$x\}$ for the sparsity (Fig.~\ref{fig:psf_memory}(a)), or the low rank-based solution (Fig.~\ref{fig:psf_memory}(b)). For both solution approaches the use of memory remains consistently at $\sim$1.068GB, which is the maximum amount of memory allocated per worker for the computations of the assigned tasks\footnote{According to the unified memory management of Spark 2.1.0, the remaining amount of memory is reserved for cached blocks immune to being evicted by execution.}. This is aligned to the iterative nature of the optimization problem at hand; during each optimization step the lineage of $\boldsymbol{\mathcal{D}}$ is updated with the new values of $\mathcal{D}_{\mathbf{X}_{p}}$, $\mathcal{D}_{\mathbf{X}_{d}}$. As a result, the length of the respective DAG and the respective number of computations undertaken by each worker increase over time. Even so, due to the adopted persistence model, the memory usage does not exceed the maximum allowable limit, and intermediate variables not fitting in memory will be recomputed each time they're needed. 

With regard to the deviations on memory usage across the different slaves we observe that four out of five slaves, allocating 4 cores on the cluster, have a similar memory usage in terms of statistical dispersion between the 25-th and 75-th percentiles, which does not exceed the value of 35MB for both approaches of optimization. Interestingly, Slave 5, which allocates 8 cores on the cluster, exhibits a greater dispersion, reaching up to $\sim$60MB for the case of sparse-based optimization and $N=$3$x$. This deviated behavior is associated with the fact that the master assigns on this slave more jobs per tasks than the remaining ones. As such, with this configuration, in order for Slave 5 to free memory space for handling more jobs, both the persistence model performs more on-the-fly computations, as well as the garbage collector cleans more frequently any unnecessary variable, thereby resulting into greater deviations (e.g. drops) on the memory usage during the execution of the experiment.

Finally, interesting remarks are derived with regard to the impact of the number $N$ of partitions per RDD on the memory usage. For both solution approaches the deviation on the memory usage is greater for $N=$3$x$ than the one recorder for $N=$6$x$. Indicatively, considering the sparse solution on Slave 5 the dispersion of the memory usage equals to 58.4MB for $N=$3$x$, opposed to 32.3MB (Fig.~\ref{fig:psf_memory}(a)), which is observed for the case of $N=$6$x$. This is due to the fact that a smaller number of partitions $N=$3$x$ results in fewer data blocks with relatively large size and increased demands in memory. This in turn stimulates the persistence model to perform more on-the-fly calculations. On the other hand, as the number of partitions increases $N=$6$x$, the size of the data blocks decreases, and subsequently, the persistence model become more relaxed. However, for the case of the sparse approach, more stages are needed to complete an action, which results into a slight increase in the execution time per loop (Fig.~\label{fig:psf_speedup}(b)). Even so, the benefit in this case is a more reliable behavior on each worker in terms of memory usage, while additionally highlighting the trade off between the RAM availability and the size of the input problem; When the memory per worker is limited with respect to size of the input data it is considered preferable to increase the number of partitions of the RDD. It may increase the time response, due to the increased number of tasks, however it will yield a more reliable solution within the lifetime of the program execution.

\begin{figure*}[htbp]
\setlength{\abovecaptionskip}{0.5pt} 
\begin{center}
\begin{tabular}{cc}
\small{(a)} & \small{(b)}\\
\includegraphics[scale=0.28]{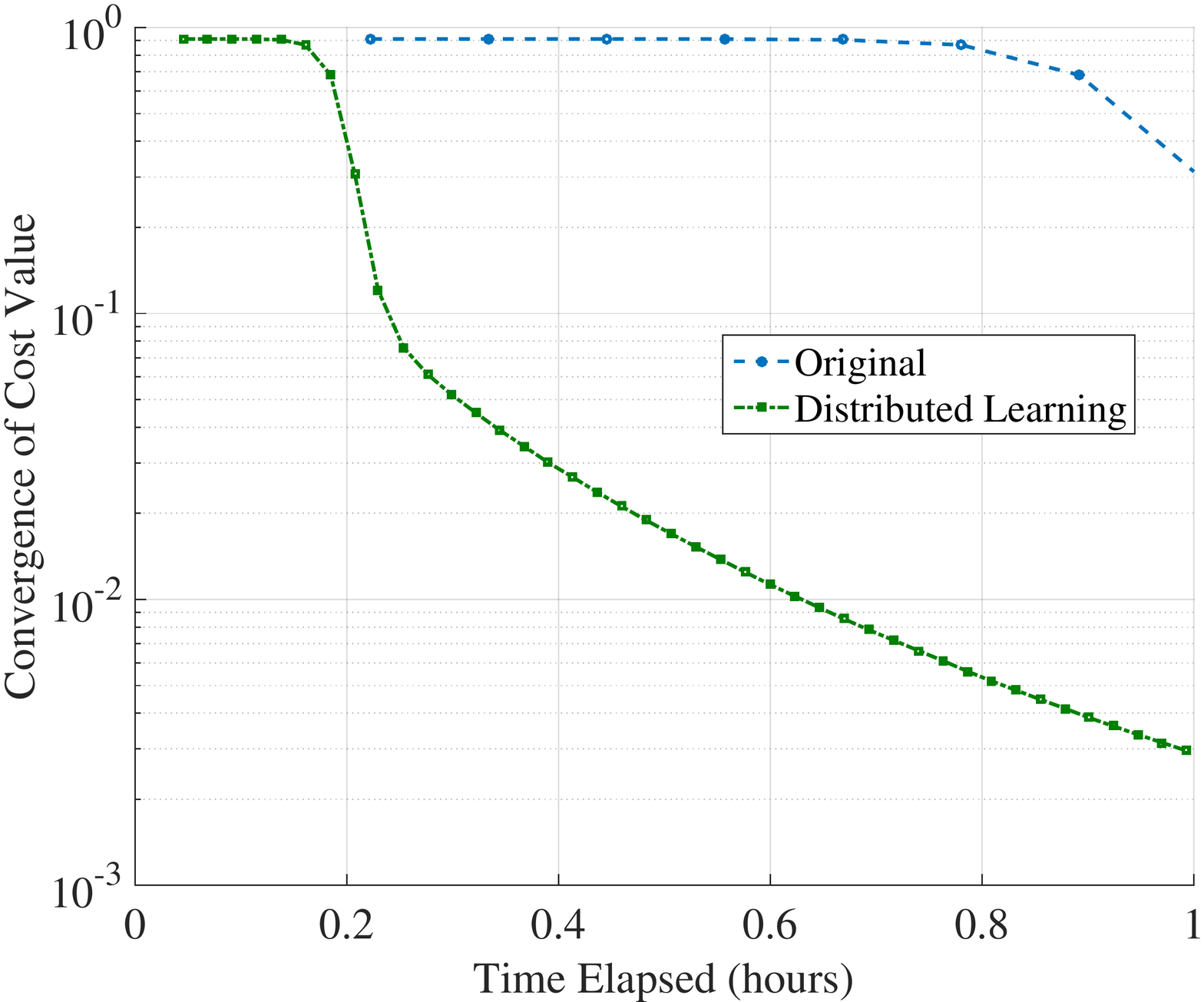} & \includegraphics[scale=0.28]{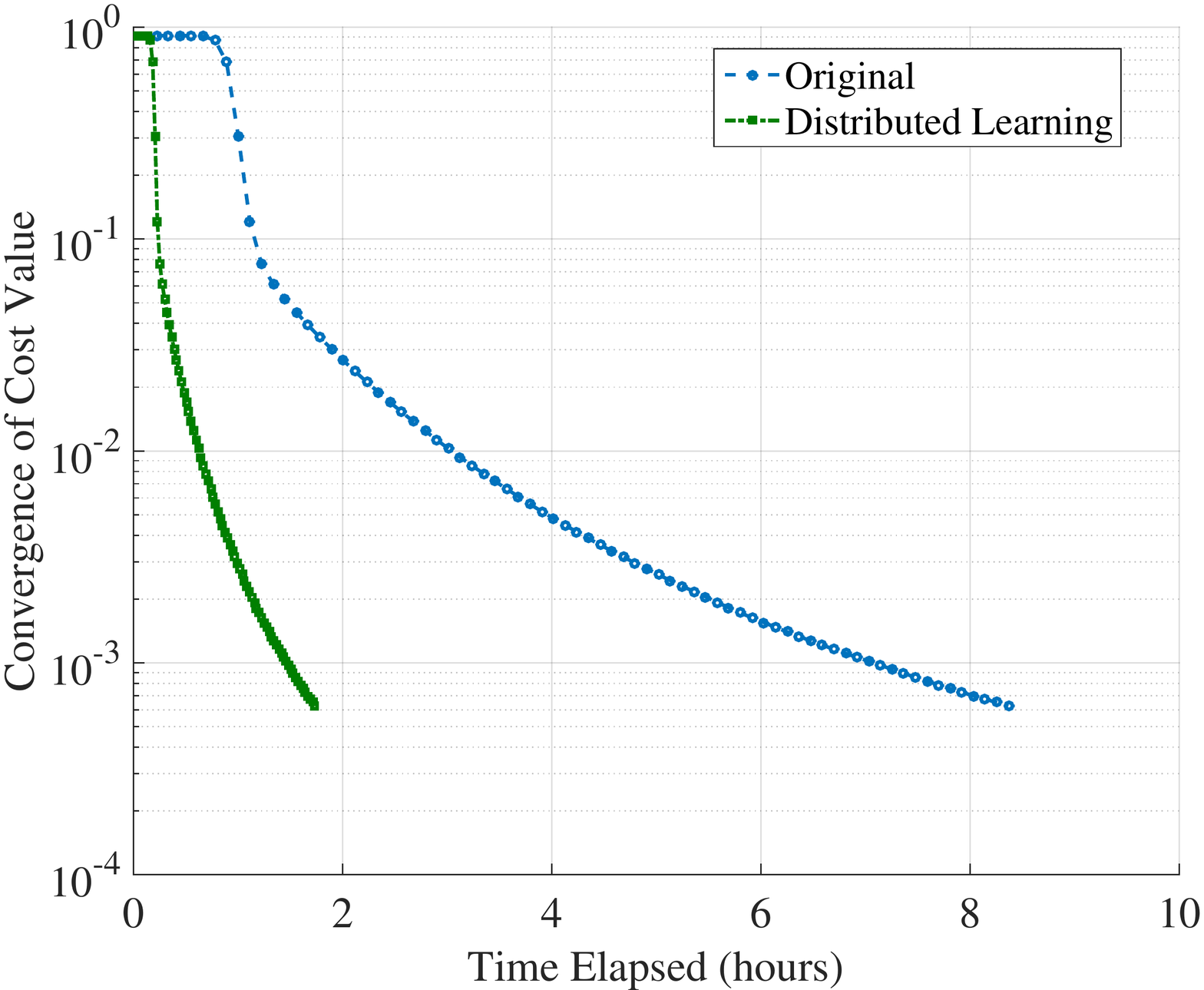}\\
\end{tabular}
\end{center}
\caption{The convergence behavior $C(\mathbf{X}_{p}^{*})$ for the spasrity-based solution, with respect to time considering the case of 20,000 images (a) 1st hour of experiment, (b) total experiment.}\label{fig:psf_convergence}
\end{figure*}

\noindent \textbf{Convergence Behavior.} Figure~\ref{fig:psf_convergence} illustrates the convergence behavior of the value of $C(\mathbf{X}_{p}^{*})$ versus the time elapsed when $i_{\max}=$ 150, and either the sequential or PSF parallelization (Algorithm~\ref{psf_parallel}, $N=$3$x$) sparsity-based approach is adopted for 20,000 images. When the proposed architecture is adopted the cost function starts converging withn the first hour of experiment, opposed to the sequential approach which can only complete a few iterations within the same time period (Fig.~\ref{fig:psf_convergence}(a)). Overall, the distributed learning approach is 75\% faster than the conventional one; the completion time of the standalone approach equals to $\sim$8 hours, opposed to the parallelized version, which does not exceed 2 hours (Fig.~\ref{fig:psf_convergence}(b)). These results highlight the fact that despite the memory overhead for storing intermediate results on each worker, the herein proposed solution is extremely beneficial in terms of time response for enabling large-scale PSF deconvolution over noisy galaxy images. 

\subsection{Remote Sensing: Super-resolution using Sparse Coupled Dictionary Training}
Current state of art in imaging systems can provide us with images from different spectral bands, giving information on the reflectivity, color, temperature or other physical properties depending on the wavelength. These images have an initial resolution ranging from a few hundred pixels (e.g., thermal infrared) to a few thousands (e.g., visible light). Even so, the highest of these resolutions is not always sufficient to distinguish small objects in a large field of view, such as in the case of an airborne carrier which looks at a distance of a few kilometers.

While single image super-resolution techniques have been developed to enhance numerically the resolution of the sensors, their main drawback is that they do not take into account structural information of the data, such as sharp borders of objects. As such, learning jointly the structure of low and high resolution images in joint dictionary learning schemes can overcome this problem and result in sharper and more visually pleasing results.

A solution to this challenge is centred around the fusion of low and high resolution training examples, in an innovativel scheme~\cite{Fotiadou2017,Fotiadou2017b} which introduces a Coupled Dictionary Learning process. Briefly, the Sparse Coupled Dictionary Learning (SCDL) algorithm formulates the super-resolution problem within the highly efficient Alternating Direction Method of Multipliers (ADMM) optimization framework~\cite{admm_conv}. As described in~\cite{Fotiadou2017,Fotiadou2017b},  this approach synthesizes a high resolution hypercube from its low resolution acquired version, by exploiting the Sparse Representations theory~\cite{Elad2010}. 
 Traditional approaches consider a set of $K$ low and high resolution image pairs and assume that these images are generated by the same statistical process under different resolution. As such, they share the same sparse coding, with respect to their corresponding low $\mathbf{X}_{l} \in \mathbb{R}^{P \times A}$, and high $\mathbf{X}_h \in \mathbb{R}^{M \times A}$ resolution dictionaries, where $A$ is the number of atoms in each dictionary.  

The coupled dictionary learning technique relies on the ADMM formulation, to yield an unconstrained version of the dictionary learning problem, which can be efficiently solved via alternating minimization. Formally, we consider the observation signals, $\mathbf{S}_{\l}=\{\mathbf{s}_{\l}\}_{i=1}^M$, and $\mathbf{S}_h=\{\mathbf{s}_{h}\}_{i=1}^P$. The main task of coupled dictionary learning is to recover both the dictionaries $\mathbf{X}_{h}$ and $\mathbf{X}_{\l}$ with their corresponding sparse codes $\mathbf{W}_{h}$ and $\mathbf{W}_{\l}$, by the means of the following $\ell_1$-minimization problem: 

\begin{eqnarray}\label{eq:scdl_eq2}
& \min_{\mathbf{X}_h,\mathbf{W}_h,\mathbf{X}_{\l},\mathbf{W}_{\l}}  ||\mathbf{S}_h-\mathbf{X}_h \mathbf{W}_h||_F^2 + ||\mathbf{S}_{\l} - \mathbf{D}_{\l} \mathbf{W}_{\l}||_F^2 +\lambda_{\l} ||\mathbf{Q}||_1 +  \lambda_h ||\mathbf{P}||_1  \\ 
& \text{s.t.}~\mathbf{P}-\mathbf{W}_h =0, \mathbf{Q}-\mathbf{W}_{\l} = 0, \mathbf{W}_h - \mathbf{W}_{\l} = 0, 
||\mathbf{D}_h(:,i)||_2 \leq 1, ||\mathbf{D}_{\l}(:,i)||_2 \leq 1 \nonumber,
\end{eqnarray}

\noindent where $\lambda_{h}$, $\lambda_{l}$ are the sparsity balance terms. The ADMM scheme considers the separate structure of each variable in Eq.(\ref{eq:scdl_eq2}), relying on the minimization of its augmented Lagrangian function: 
\begin{align}\label{eq:scdl_eq3}
& L(\mathbf{X}_h,\mathbf{X}_{\l},\mathbf{W}_h,\mathbf{W}_{\l},\mathbf{P},\mathbf{Q},\mathbf{Y}_1,\mathbf{Y}_2,\mathbf{Y}_3) \notag  = ||\mathbf{X}_h \mathbf{W}_h -  \mathbf{S}_h||_F^2 + ||\mathbf{X}_{\ell} \mathbf{W}_{\ell} - \mathbf{S}_{\ell}||_F^2 + \lambda_h ||\mathbf{P}||_1 \notag \\ & + \lambda_{\ell} ||\mathbf{Q}||_1  +  <Y_1,\mathbf{P} - \mathbf{W}_h>  
 + <Y_2, \mathbf{Q} - \mathbf{W}_{\ell}> +  <Y_3,\mathbf{W}_h - \mathbf{W}_{\ell}>  +  \frac{c_1}{2} ||\mathbf{P} - \mathbf{W}_h||_F^2 
\notag \\ & +  \frac{c_2}{2} ||\mathbf{Q} - \mathbf{W}_{\ell}||_F^2 + \frac{c_3}{2} ||\mathbf{W}_h - \mathbf{W}_{\ell}||_F^2,
\end{align}

\noindent where ${\bf Y}_1$, ${\bf Y}_2$, and ${\bf Y}_3$ stand for the Lagrange multiplier matrices, while $c_1>0$, $c_2>0$, and $c_3>0$ are the respective the step size parameters.  In each optimization step, the updated dictionary matrices are the following~(Algorithm 1 in~\cite{Fotiadou2017b}):

\begin{eqnarray}
\mathbf{X}_h &=&  \mathbf{X}_h + \frac{\mathbf{S}_{h} \times \mathbf{W}_{h}}{\phi_{h} + \delta} \label{eq:scdl_eq3}\\ 
\mathbf{X}_l &=&  \mathbf{X}_l + \frac{\mathbf{S}_{l} \times \mathbf{W}_{l}}{\phi_{l} + \delta} \label{eq:scdl_eq4},
\end{eqnarray}

\noindent where $\phi_{h} = \mathbf{W}_{h} \times \mathbf{W}_{h}^{T}$, $\phi_{l} = \mathbf{W}_{l} \times \mathbf{W}_{l}^{T}$, and $\delta$ is the regularization factor.
 
As explained in~\cite{Fotiadou2017, Fotiadou2017b} Sparse  Representations and Coupled  Dictionary Learning are powerful tools for reconstructing spectral profiles from their corresponding low-resolution, and noisy  versions, with tolerance in extreme noise  scenarios.
\subsubsection{Parallelization using the distributed learning architecture}
\begin{figure*}[htbp]
\begin{center}
\includegraphics[scale=0.2]{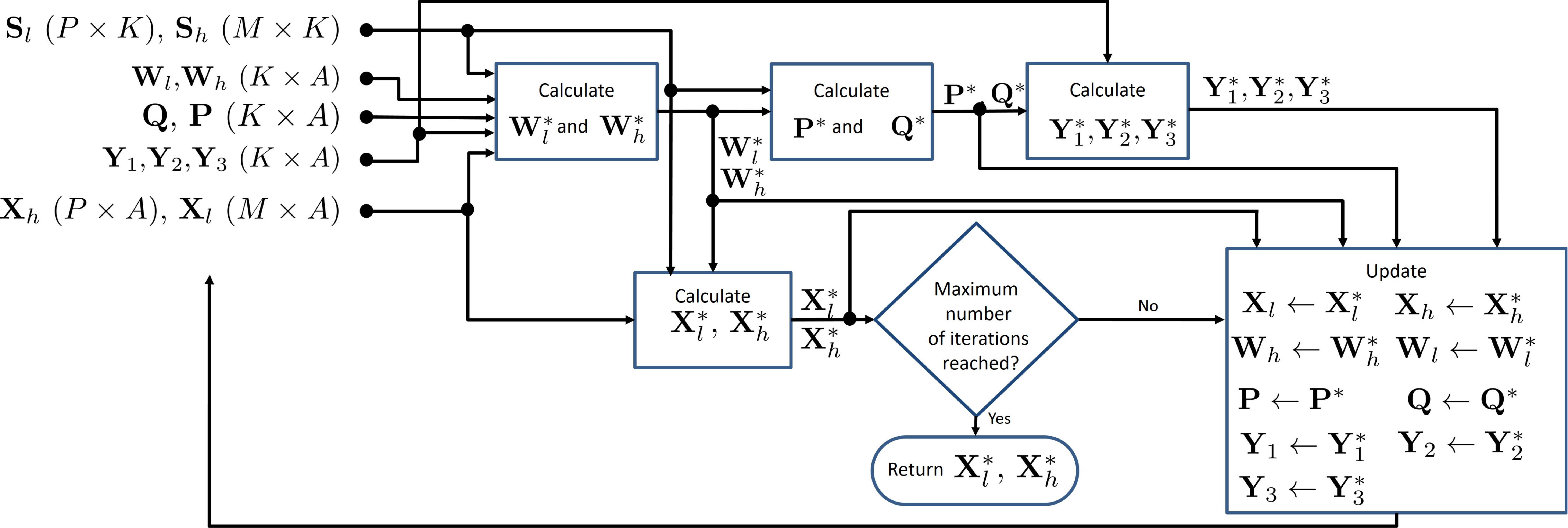}
\caption{The flow diagram for the sequential implementation of the sparse coupled dictionary learning algorithm.}
\label{fig:cdl_flow}
\end{center}
\end{figure*}

The general algorithmic strategy of ADMM scheme for calculating $\mathbf{X}_{\l}$ and $\mathbf{X}_{h}$ seeks a stationary point by solving for one of the variables, while keeping the others fixed.  The sequential approach for solving Eq.~\ref{eq:scdl_eq3} reflects on an iterative process. As elaborated in~\cite{Fotiadou2017, Fotiadou2017b}, during each iteration step all variables involved (i.e., $\mathbf{S}_{h}$, $\mathbf{S}_{l}$, $\mathbf{W}_{h}$, $\mathbf{W}_{l}$, $\mathbf{P}$, $\mathbf{Q}$, $\mathbf{Y}_{1,2,3}$) need to be jointly processed in matrix operations for updating the values of $\mathbf{X}_{h}$ and $\mathbf{X}_{l}$  (Fig~\ref{fig:cdl_flow}). 

Both the flow diagram presented in Fig.~\ref{fig:cdl_flow} as well as the calculation steps explained in~\cite{Fotiadou2017b} highlight the importance of the intermediate sparse coding and Laplacian multiplier matrices for calculating the dictionaries. Notably these intermediate matrices have size $K \times A$ and as such, the sequential approach could readily become inefficient as the number of data samples $K$ and the combination of $P$, $M$ and $A$ increases. 

\begin{algorithm*}[ht!]
\caption{The SCDL algorithm parallelization (\colorbox{red!20}{steps} are performed on the driver, while \colorbox{green!10}{steps} are performed over the cluster).}
\label{scdl_parallel}
\begin{footnotesize}
\KwData{The high $\mathbf{S}_{h}$ ($\in \mathbb{R}^{P \times K}$) and the low $\mathbf{S}_{l}$ ($\in \mathbb{R}^{M \times K}$) 3D cubes, the number of dictionary atoms $A$, and the maximum number of iterations $i_{\max}$. Typically: $i_{\max}=$100.}
\KwResult{The high $\mathbf{X}_{h}^{*}$ ($\in \mathbb{R}^{P \times A}$) and low $\mathbf{X}_{l}^{*}$ ($\in \mathbb{R}^{M \times A}$) resolution dictionary matrices.}
\nl \HiLiDR Define the RDD for $\mathbf{S}_{l}$ and $\mathbf{S}_{h}$ as $\mathcal{D}_{\mathbf{X}_{h}}$, $\mathcal{D}_{\mathbf{X}_{l}}$ respectively, with $N$ partitions per RDD.\\
\nl \HiLiCL Initialize $\mathbf{X}_{h}$ and $\mathbf{X}_{l}$, considering $P \times A$ and $M \times K$ random samples from $\mathcal{D}_{\mathbf{X}_{h}}$ and $\mathcal{D}_{\mathbf{X}_{l}}$ respectively. \\
\nl \HiLiCL Return $\mathbf{X}_{h}$, $\mathbf{X}_{l}$ to driver.\\
\nl \HiLiCL Create the RDD bundle: $\boldsymbol{\mathcal{D}} = \mathcal{D}_{\mathbf{X}_{h}}.\text{\texttt{zip}}(\mathcal{D}_{\mathbf{X}_{l}})$.\\
\nl \HiLiCL Map $\boldsymbol{\mathbf{D}}$ into $\mathcal{D}_{\mathbf{Z}}$ for $\mathbf{Z} = \{\mathbf{W}_{h}, \mathbf{W}_{l}, \mathbf{P}, \mathbf{Q}, \mathbf{Y}_{1},\mathbf{Y}_{2},\mathbf{Y}_{3}\}$, by creating a RDD SCDL object $\mathcal{D}_{SCDL}$, containing\\ \HiLiCL $N$ blocked parallel partitions for all intermediate matrices $\mathbf{Z}$, i.e., $\mathcal{D}_{SCDL} \triangleq [\mathcal{D}_{\mathbf{X}_{h}}, \mathcal{D}_{\mathbf{X}_{l}}, \mathcal{D}_{\mathbf{W}_{h}}, \mathcal{D}_{\mathbf{X}_{l}}, \mathcal{D}_{\mathbf{P}}, $ \\ \HiLiCL $\mathcal{D}_{\mathbf{Q}}, \mathcal{D}_{\mathbf{Y}_{1}},\mathcal{D}_{\mathbf{Y}_{2}}, \mathcal{D}_{\mathbf{Y}_{3}}]$, : $\mathcal{D}_{SDCL} = \boldsymbol{\mathcal{D}}.\text{\texttt{map}}(\text{\texttt{lambda} } x: $ $ \text{Start~SCDL~}(x, A)))$.\\
\nl \HiLiDR\For{$i:1:i_{\max}$}{ 
\nl \HiLiDR Broadcast $\mathbf{X}_{h}$ and $\mathbf{X}_{l}$, and respective auxiliary transposed and inversed variables~\cite{Fotiadou2017b}.\\
\nl \HiLiCL Update $\mathcal{D}_{\mathbf{Z}}$, for $\mathbf{Z} = \{\mathbf{W}_{h}, \mathbf{W}_{l}, \mathbf{P}, \mathbf{Q}, \mathbf{Y}_{1},\mathbf{Y}_{2},\mathbf{Y}_{3}\}$ in $\mathcal{D}_{SCDL}$ using Alg. 1,~\cite{Fotiadou2017b}:  \\ \HiLiCL $\mathcal{D}_{CSDL}=\mathcal{D}_{CSDL}.\text{\texttt{map}}(\text{\texttt{lambda} }x: \text{Update } x)$.\\
\nl \HiLiCL Calculate the auxiliary variables $\mathbf{X}_{h} \times \mathbf{W}_{h}$, $\mathbf{X}_{l} \times \mathbf{W}_{l}$, $\phi_{h}=\mathbf{W}_{h} \times \mathbf{W}_{h}^{T}$, $\phi_{l}=\mathbf{W}_{l} \times \mathbf{W}_{l}^{T}$ (Alg. 1,~\cite{Fotiadou2017b}): \\ \HiLiCL $[\mathbf{X}_{h} \times \mathbf{W}_{h}, \mathbf{X}_{l} \times \mathbf{W}_{l}, \phi_{h}, \phi_{l}]= \mathcal{D}_{SCDL}.\text{\texttt{map}}(\text{\texttt{lambda} } x:$ $\text{Calc. Outer Product}(x)).\text{\texttt{reduce}(\texttt{lambda} }x,y: $ \\ \HiLiCL $\text{Calc. Outer Product}(x,y))$.\\
\nl \HiLiDR Update the dictionary matrices $\mathbf{X}_{h}$,  $\mathbf{X}_{l}$ according to Eq.~(\ref{eq:scdl_eq3})-(\ref{eq:scdl_eq4}).\\
}
\nl \HiLiDR Save final dictionary matrices $\mathbf{X}_{h}^{*} \leftarrow \mathbf{X}_{h}$,  $\mathbf{X}_{l}^{*} \leftarrow \mathbf{X}_{l}$.\\
\end{footnotesize}
\vspace*{-0.05cm}
\end{algorithm*}
\setlength{\textfloatsep}{0pt}

The parallelization of this scheme is described in Algorithm~\ref{scdl_parallel} for distributing the essential matrix calculations, according to the proposed distributed learning architecture. In compliance to parallel efforts (e.g.,~\cite{Makkie2018}) we assume that dictionaries fit into the memory of a single computer. The first step considers the parallelization of the input imaging 3D cubes $\mathbf{S}_{h}$, $\mathbf{S}_{l}$ into $\mathcal{D}_{\mathbf{S}_{h}}$, $\mathcal{D}_{\mathbf{S}_{l}}$, on the side of the driver program and their compression into the $\boldsymbol{\mathcal{D}}$, which essentially contains tuples of the form $<\mathbf{S}_{h}, \mathbf{S}_{l}>$ (i.e., $k=$2). The initial values of the dictionaries $\mathbf{X}_{h}$, $\mathbf{X}_{l}$ contain random samples of $\mathcal{D}_{\mathbf{S}_{h}}$, $\mathcal{D}_{\mathbf{S}_{l}}$ collected over the cluster at the driver program. The RDD bundle is $\boldsymbol{\mathcal{D}}$ is transformed into a SCDL object $\mathcal{D}_{SCDL}$ which enriches $\boldsymbol{\mathcal{D}}$ with the parallelized version of all auxiliary matrices i.e. $\mathcal{D}_{\mathbf{W}_{h}}, \mathcal{D}_{\mathbf{X}_{l}}, \mathcal{D}_{\mathbf{P}}, \mathcal{D}_{\mathbf{Q}}, \mathcal{D}_{\mathbf{Y}_{1}},\mathcal{D}_{\mathbf{Y}_{2}}, \mathcal{D}_{\mathbf{Y}_{3}}$. During each iteration, the driver broadcasts the current dictionaries $\mathbf{X}_{h}$, $\mathbf{X}_{l}$ to the cluster, which are in turn employed for updating the auxiliary matrices contained in $\mathcal{D}_{SCDL}$ in a distributed fashion,  according to Algorithm~1 in~\cite{Fotiadou2017b}. By exploiting the properties of outer products calculations, the firing of the respective combination of map-reduce action over $\mathcal{D}_{SCDL}$ (Step 9 in Algorithm~\ref{scdl_parallel}) triggers the calculation of auxiliary structures $\mathbf{S}_{h} \times \mathbf{W}_{h}$,  $\mathbf{S}_{h} \times \mathbf{W}_{h}$, and $\phi_{h}$, $\phi_{l}$, which are essential for updating the dictionary matrices $\mathbf{X}_{h}$, $\mathbf{X}_{l}$ in Eq.(~\ref{eq:scdl_eq3}), Eq.(~\ref{eq:scdl_eq4}), respectively. This process is repeated until the maximum number of iterations is reached, while the final dictionaries $\mathbf{X}_{h}^{*}$, $\mathbf{X}_{l}^{*}$ are directly saved on the memory of the driver program.

\subsubsection{Evaluation Studies}\label{sec:csdl_studies}
Similarly to the case of the space variant deconvolution, the objective the benchmark studies is to evaluate the performance of the SCDL algorithm parallelization implemented over the herein proposed learning architecture. Two types of datasets are herein considered, namely: (a) hyperspectral (HS) data, comprised of video streaming frames on different spectral bands, captured by snapshot mosaic sensors that feature IMEC Spectrally Resolvable Detector Array~\cite{Lambrechts} ($P=$5$\times$5, $M=$3$\times$3, $K\backsimeq$40,000), and (b) grayscale (GS) data  ($P=$17$\times$17, $M=$9$\times$9, $K\backsimeq$40,000), extracted from the Berkeley Segmentation Dataset~\cite{fowlkes2014berkeley}.

The distributed learning architecture retains the same characteristics of the private cluster described in Section~\ref{sec:psf_studies}, i.e. Apache Spark 2.1.0. 2017 release over 5 slaves. In order to better accommodate the computational needs of this use case, we configure one of the slaves to generate 2 Spark workers, each allocating 2.8GB RAM and 4 CPU cores. As a result, the cluster yields in total 24 CPU cores and 16.8GB RAM. With regard to the data persistence model, we consider both the memory only, as well as the memory-and-disk model, according to which RDD data not fitting in memory are instead stored on disk and read from there when they are needed.

Similarly to the astrophysics use case (Section~\ref{psf_usecase}), the evaluation procedure emphasizes on the speedup with respect to the sequential implementation\footnote{\href{https://github.com/spl-icsforth/SparseCoupledDictionaryLearning}{\textcolor{blue}{https://github.com/spl-icsforth/SparseCoupledDictionaryLearning}}}, the memory and disk usage, and the convergence rate of the normalized root mean square error in both low and high resolution. The key experimental parameters are the type of input data (HS or GS) and respective parameters (i.e., $M$, $P$), and the dictionary size $A$ with respect to the number of partitions $N$ partitions per RDD. We herein consider $N=\{2x, 3x, 4x, 6x\}$, where $x$ corresponds to the total number of cores available for parallel calculations, i.e. $x=$24. 

\begin{figure*}[h!]
\setlength{\abovecaptionskip}{0.5pt} 
\begin{center}
\begin{tabular}{cc}
\small{(a)} & \small{(b)}\\
\includegraphics[scale=0.28]{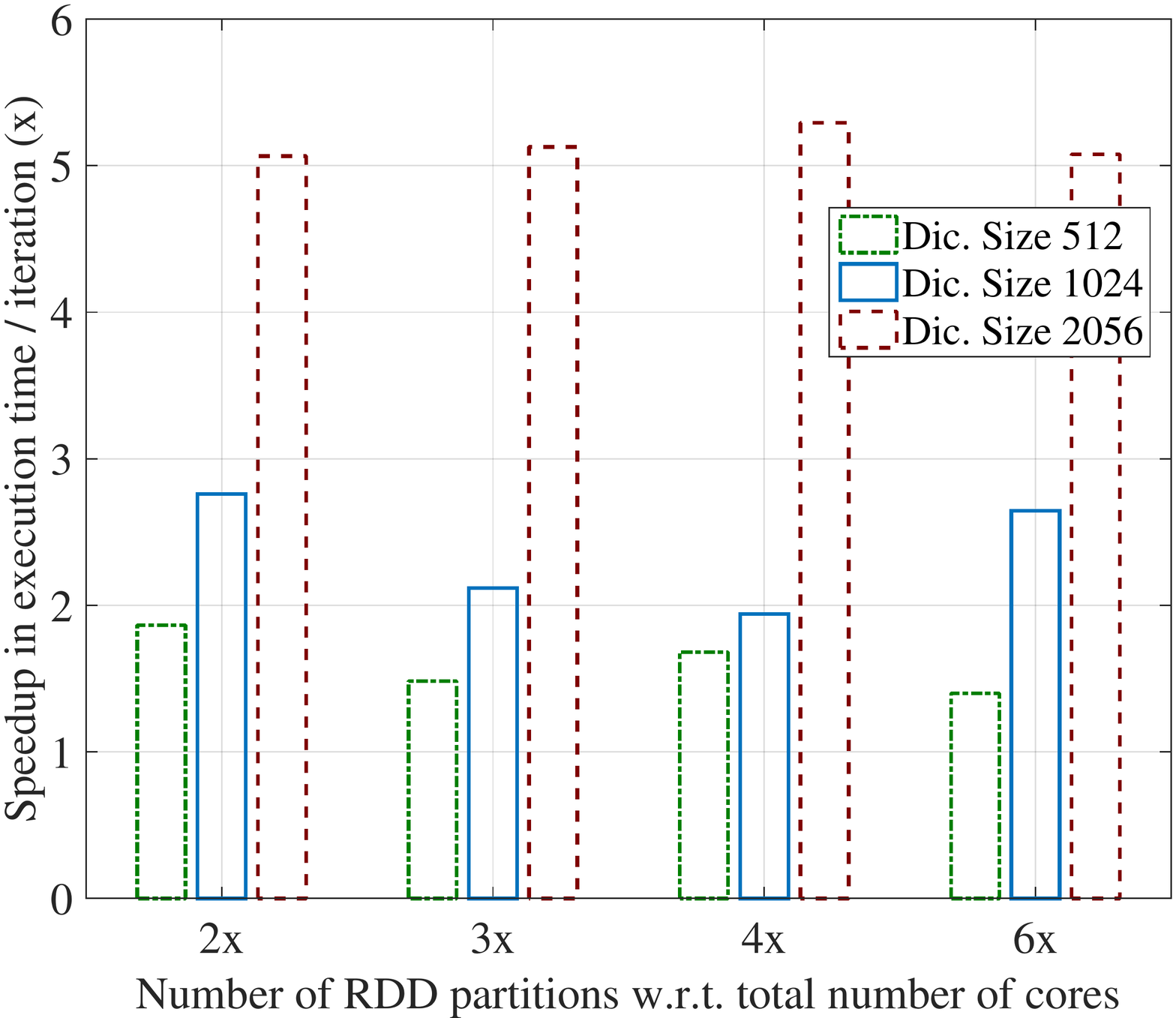} & \includegraphics[scale=0.28]{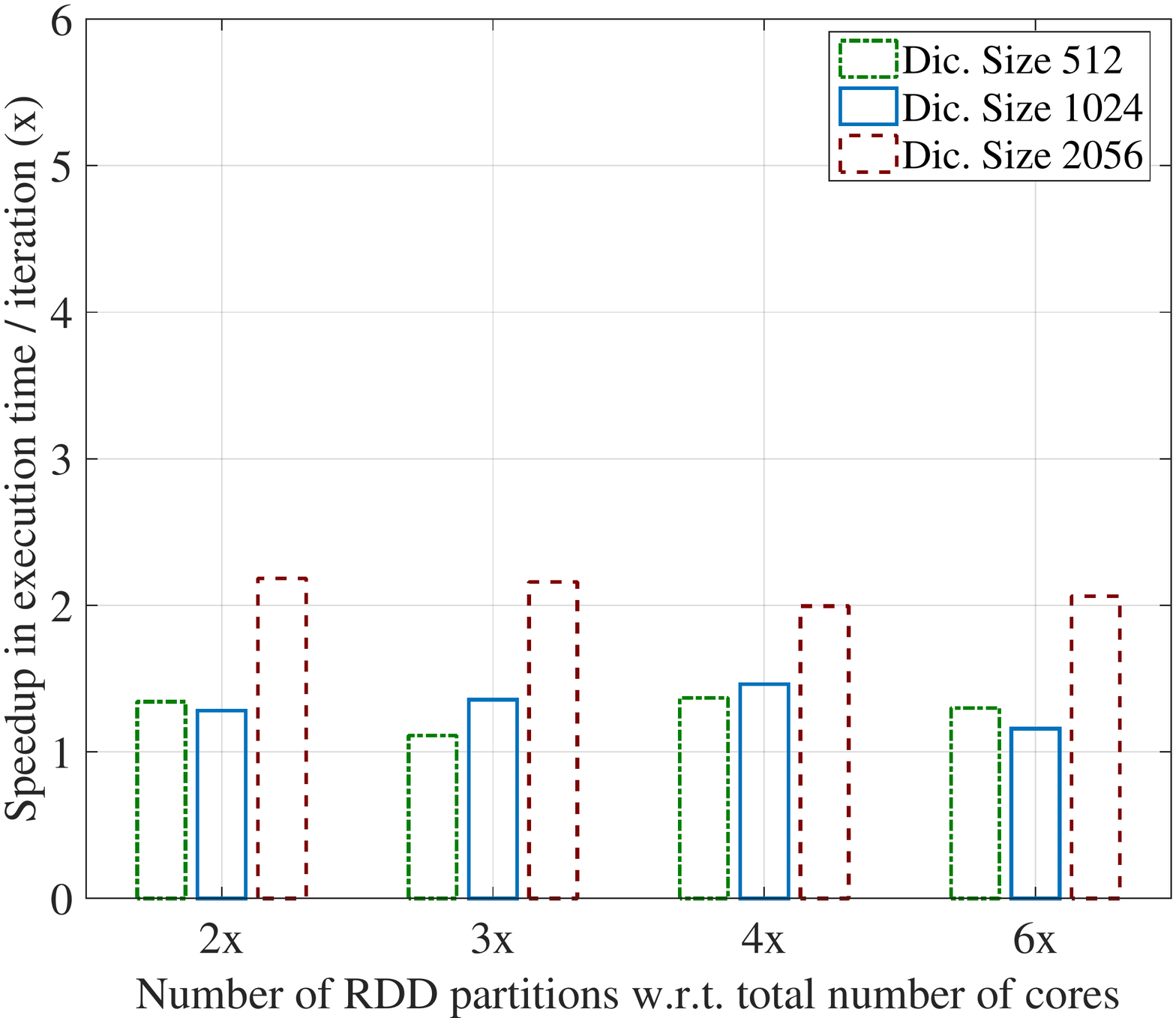}\\
\small{(c)} & \small{(d)}\\
\includegraphics[scale=0.28]{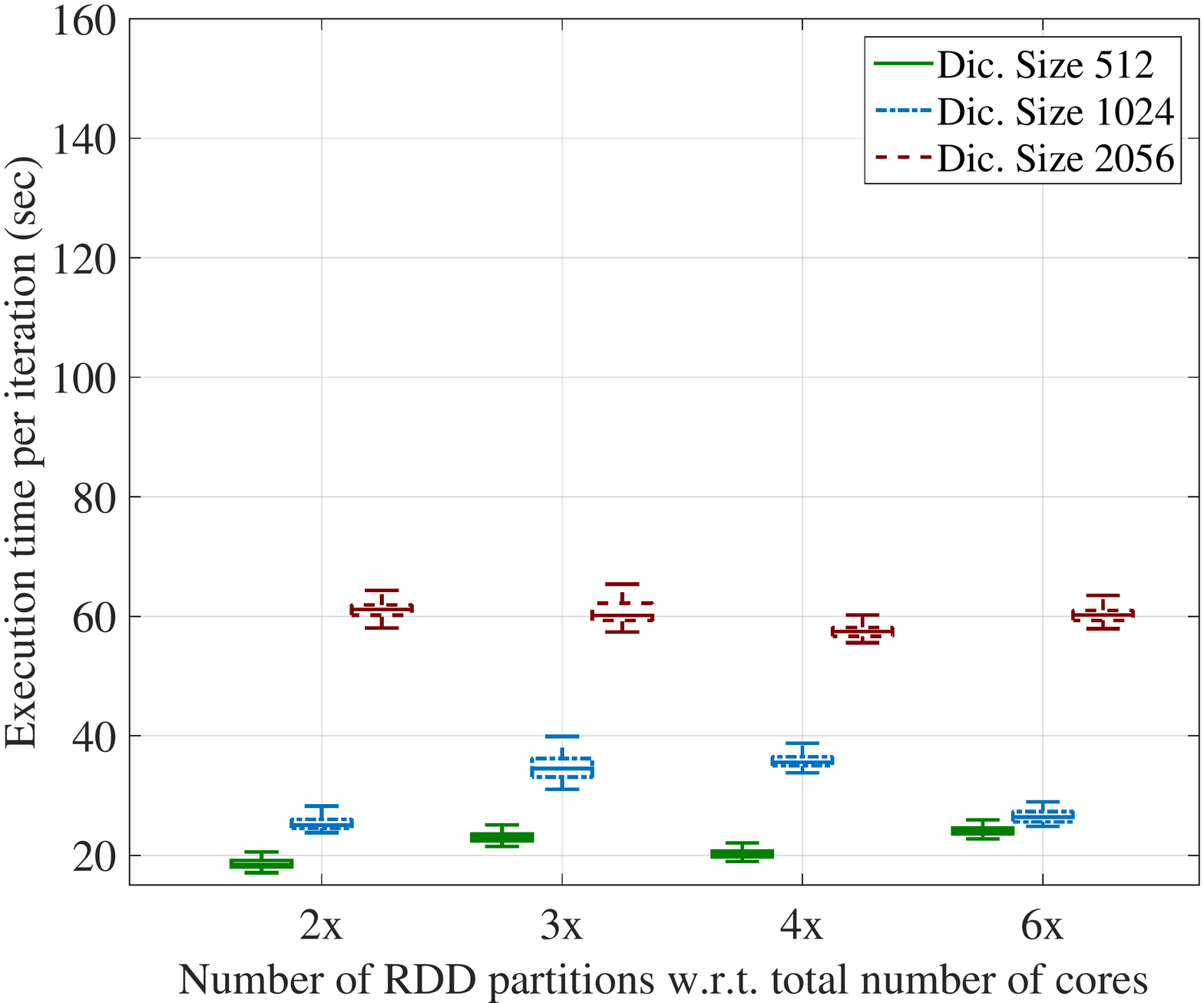} & \includegraphics[scale=0.28]{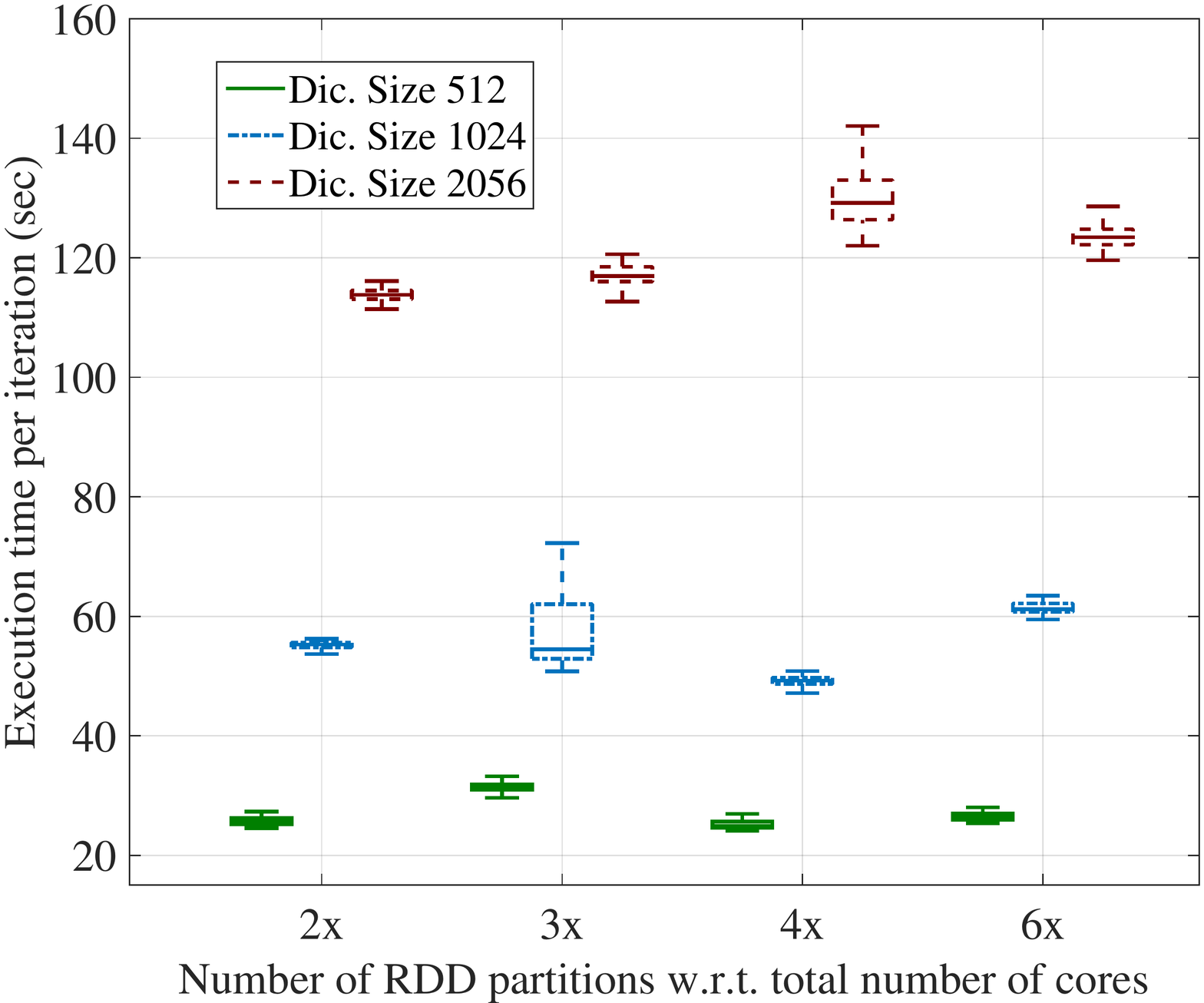}\\
\end{tabular}
\end{center}
\caption{Time performance for the parallelization of the SCDL-based super resolution based on the proposed architecture: (a) speedup for the HS imaging data, (b) speedup for the GS imaging data, (c) time execution per optimization loop for the HS data, (d) time execution per optimization loop for the GS data.}\label{fig:cdl_speedup}
\end{figure*}

\noindent\textbf{Speedup.} Figure~\ref{fig:cdl_speedup} presents the time performance achieved for both HS as well as GS imaging data in terms of speedup (Fig.~\ref{fig:cdl_speedup}(a), Fig.~\ref{fig:cdl_speedup}(b)) and execution time per iteration (Fig.~\ref{fig:cdl_speedup}(c), Fig.~\ref{fig:cdl_speedup}(d)) for the calculation of $\mathbf{X}_{h}^{*}$, and $\mathbf{X}_{l}^{*}$. Increased speed up is offered for both for the HS as well as the GS data, as the number of dictionary atoms ($D$) increases for all cases of $N$. Specifically, for the HS data (Fig.~\ref{fig:cdl_speedup}(a)) the speedup increases $\sim$2.5$x \rightarrow$ 5$x$ when $D$ increases 1024 to 2056. The impact of the number of partitions $N$ per RDD on the time performance is more evident for lower values of $D$; for example for $D=$512, the optimal speedup is achieved for $N=$2$x$, and decreases as the number of RDD blocks increases. This is consistent to the PSF use case (10,000 images), and highlights that when the size of the tasks fits in the memory of each worker, it is more beneficial to retain a small value of $N$, in order to avoid the unnecessary Spark network and computational overheads. Interestingly, as the size $D$ of the dictionaries increases speedup becomes less variant on the number of partitions; for $D=$2056 the speedup achieved is $\sim$ 5$x$ for all cases of $N$ examined. Similar remarks can be derived for the GS data (Fig.~\ref{fig:cdl_speedup}(b)),wherein we additionally observe that due to increased size of the problem ($P=$17$\times$17, $M=$9$\times$9) the speedup values offered are smaller than the ones provided for the HS data. Nevertheless, when $D=$2056, the proposed scheme yields speedup that remains higher than 2$x$.  

With respect to the time execution per optimization loop (Fig.~\ref{fig:cdl_speedup}(c) and Fig.~\ref{fig:cdl_speedup}(d)) the distributed architecture yields a stable performance, since for all combinations of data type, dictionary size, and value of $N$ the time execution has limited statistical dispersion between the 25-th and 75-th percentiles. As expected, an increase to the size $D$ of the dictionaries leads to an increase to the execution time, since the size of the problem significantly changes. The impact of the parameter $N$ on the time performance is more evident for the case of the GS data and $D=$2056, indicating that for this problem, it is more beneficial to retain a smaller number of RDD partitions. Indicatively, when $N=2$x the median of execution time remains below 120secs, opposed to case of $N=$4$x$ wherein the execution time per iteration reaches 130secs. This is consistent to the speedup results, and suggest that the increase of $N$ implies more data chunks of smaller size need to be exchanged among the workers, thereby introducing unnecessary shuffling overhead. By contrast, when the value of $N$ is retained lower, the partitioning into less data chunks compensates such phenomena. 

\begin{figure}[htbp]
\setlength{\abovecaptionskip}{0.5pt} 
\begin{center}
\includegraphics[scale=0.3]{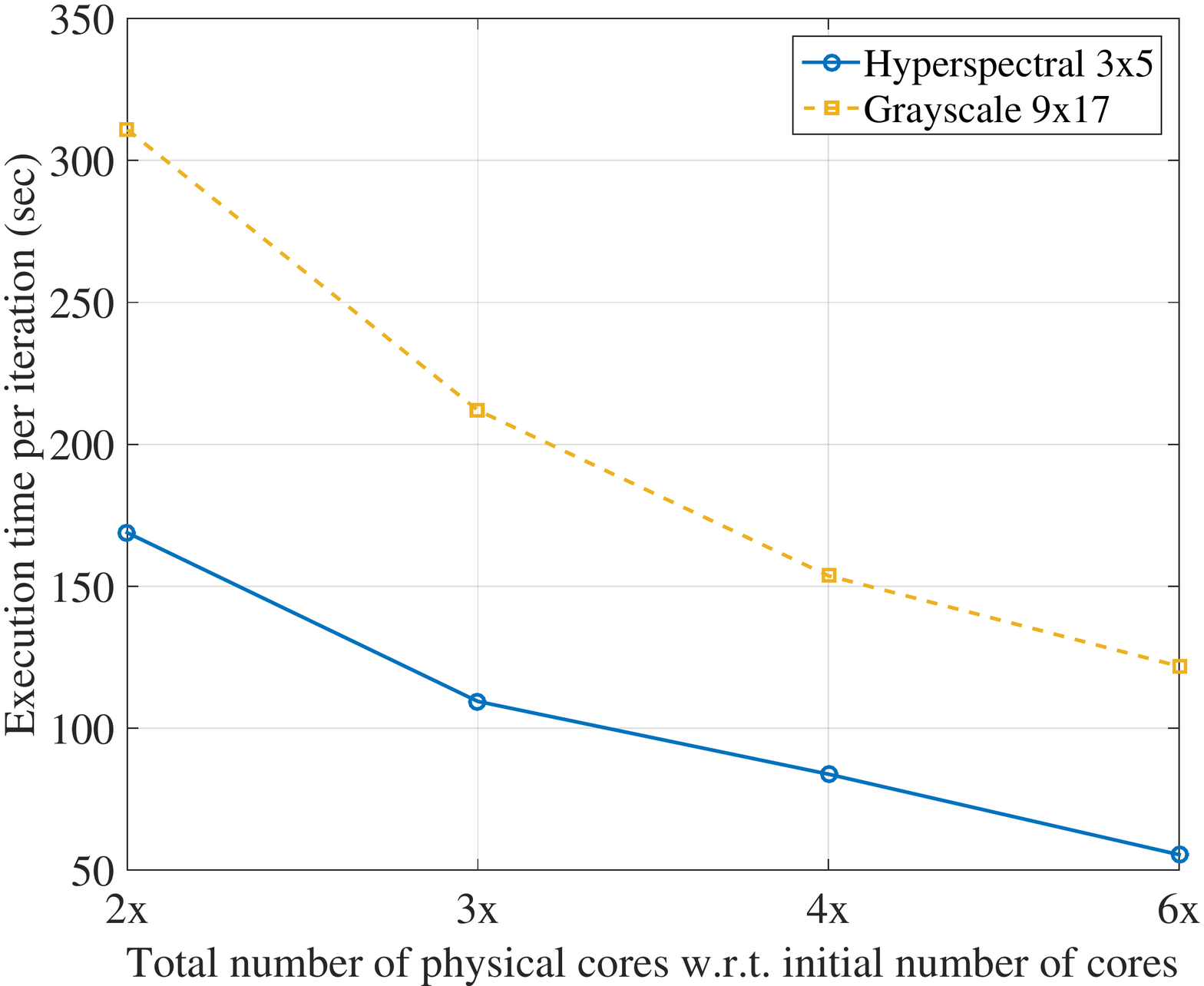}
\end{center}
\caption{Scalability of execution time per iteration for the HS and GS imaging data.}\label{fig:cdl_scalability}
\end{figure}
 
Scalability aspects on deploying Algorithm~\ref{scdl_parallel} over the cluster infrastructure are illustrated in Fig~\ref{fig:cdl_scalability} for the $D=$2056 and $N=$ 4$x$, for both HS and GS data. As expected, as the total number of cores increases with respect to the cores considered for the evaluation of the sequential approach (i.e., 4 cores), the distributed learning approach offers substantial improvements in terms of time performance. Specifically, increasing the number of available cores in the cluster from 2$x$ (i.e., 8 cores) to 6$x$ (i.e., 24 cores) results into a 61.3\% (310sec $\rightarrow$ 120sec) and 72.2\% (180sec $\rightarrow$ 50sec) improvement for the GS and HS data, respectively. 

\begin{figure}[htbp]
\setlength{\abovecaptionskip}{0.5pt} 
\begin{center}
\includegraphics[scale=0.3]{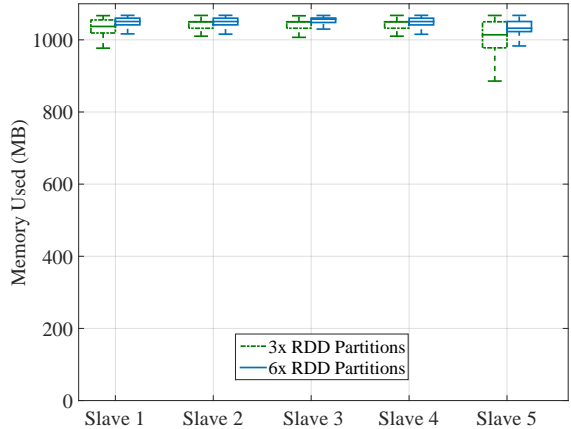}\\
\end{center}
\caption{Memory usage per slave for the HS data}.\label{fig:cdl_hsmemory}
\end{figure}

\begin{figure*}[h!]
\setlength{\abovecaptionskip}{0.5pt} 
\begin{center}
\begin{tabular}{cc}
\small{(a)} & \small{(b)}\\
\includegraphics[scale=0.28]{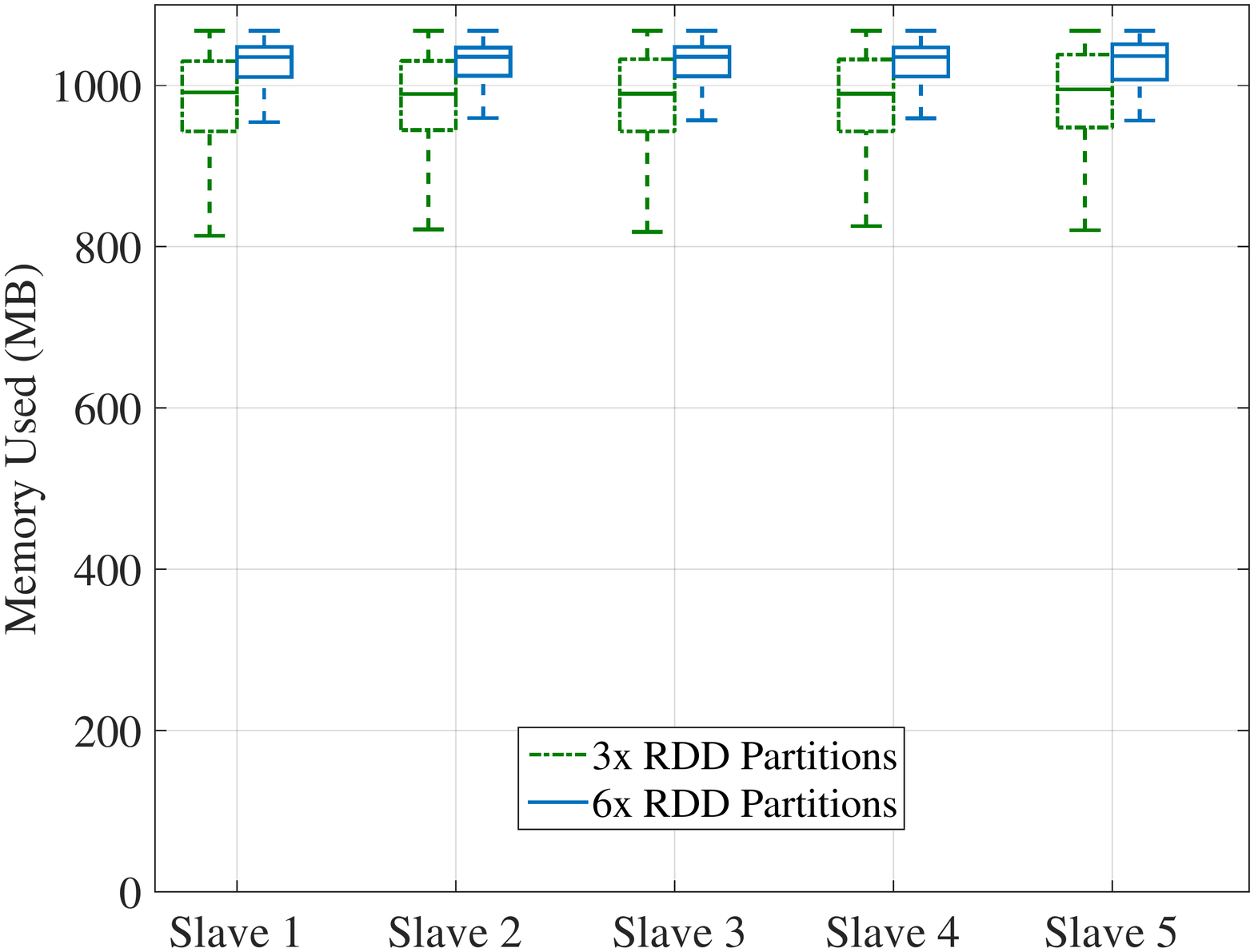} & \includegraphics[scale=0.28]{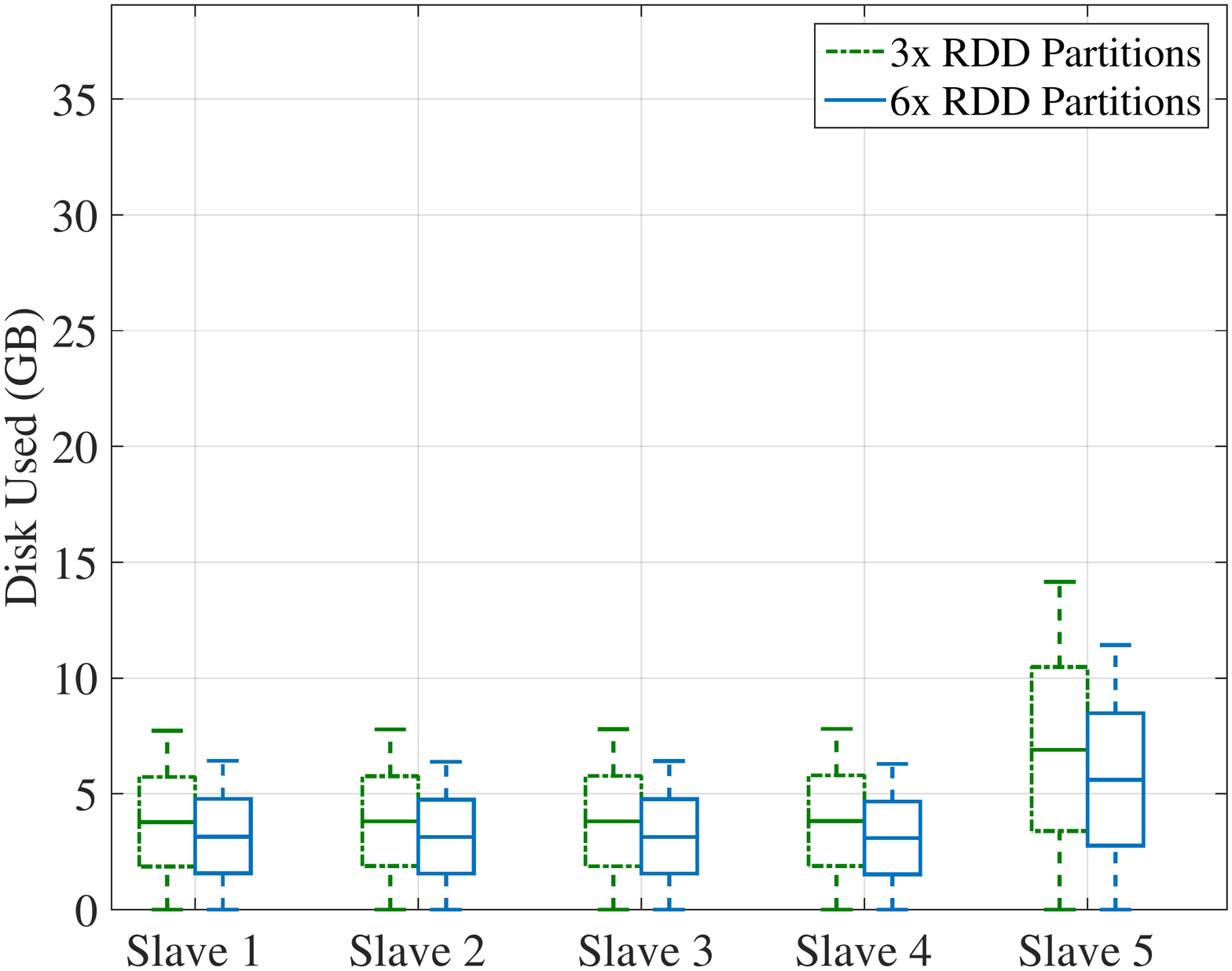}\\
\small{(c)} & \small{(d)}\\
\includegraphics[scale=0.28]{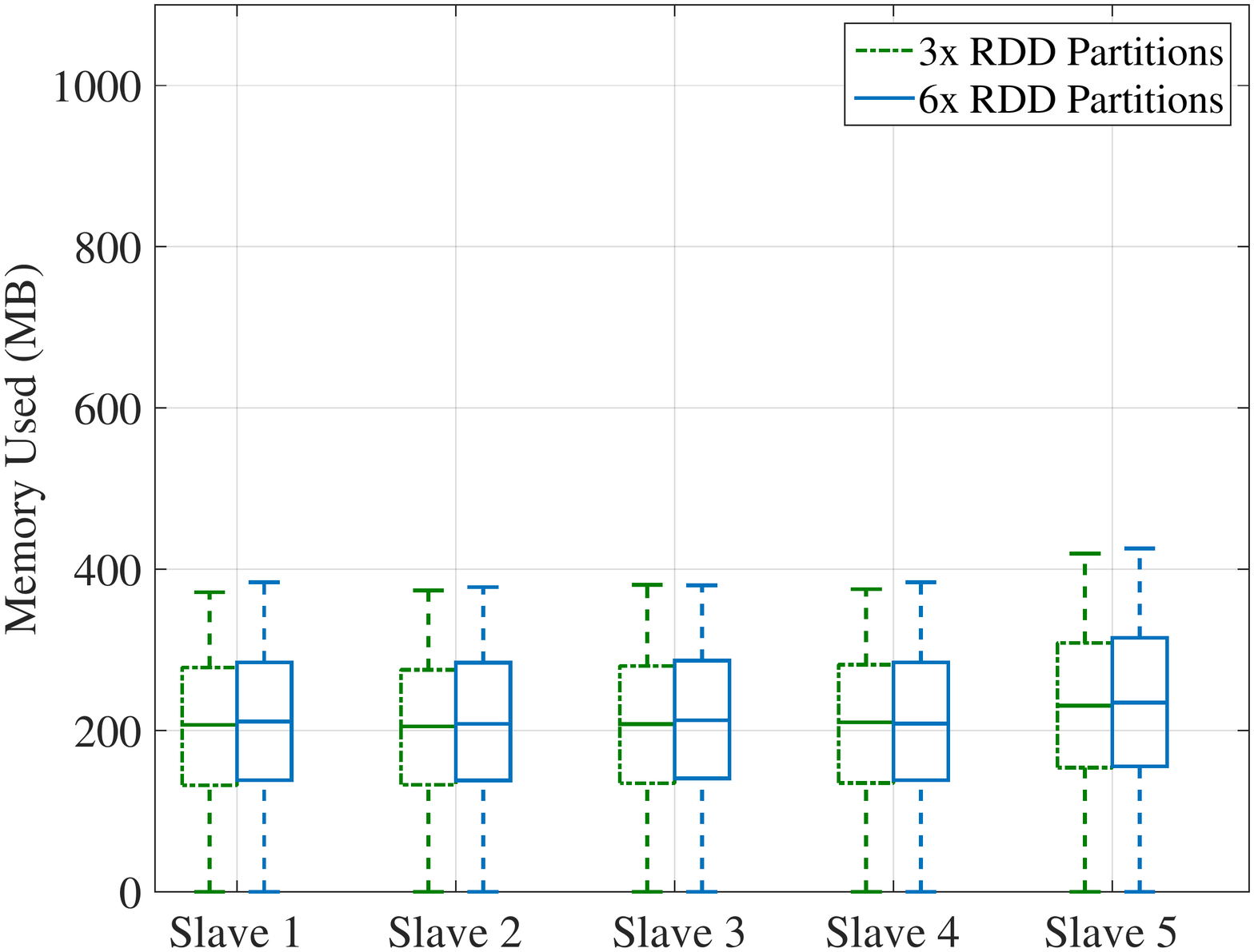} & \includegraphics[scale=0.28]{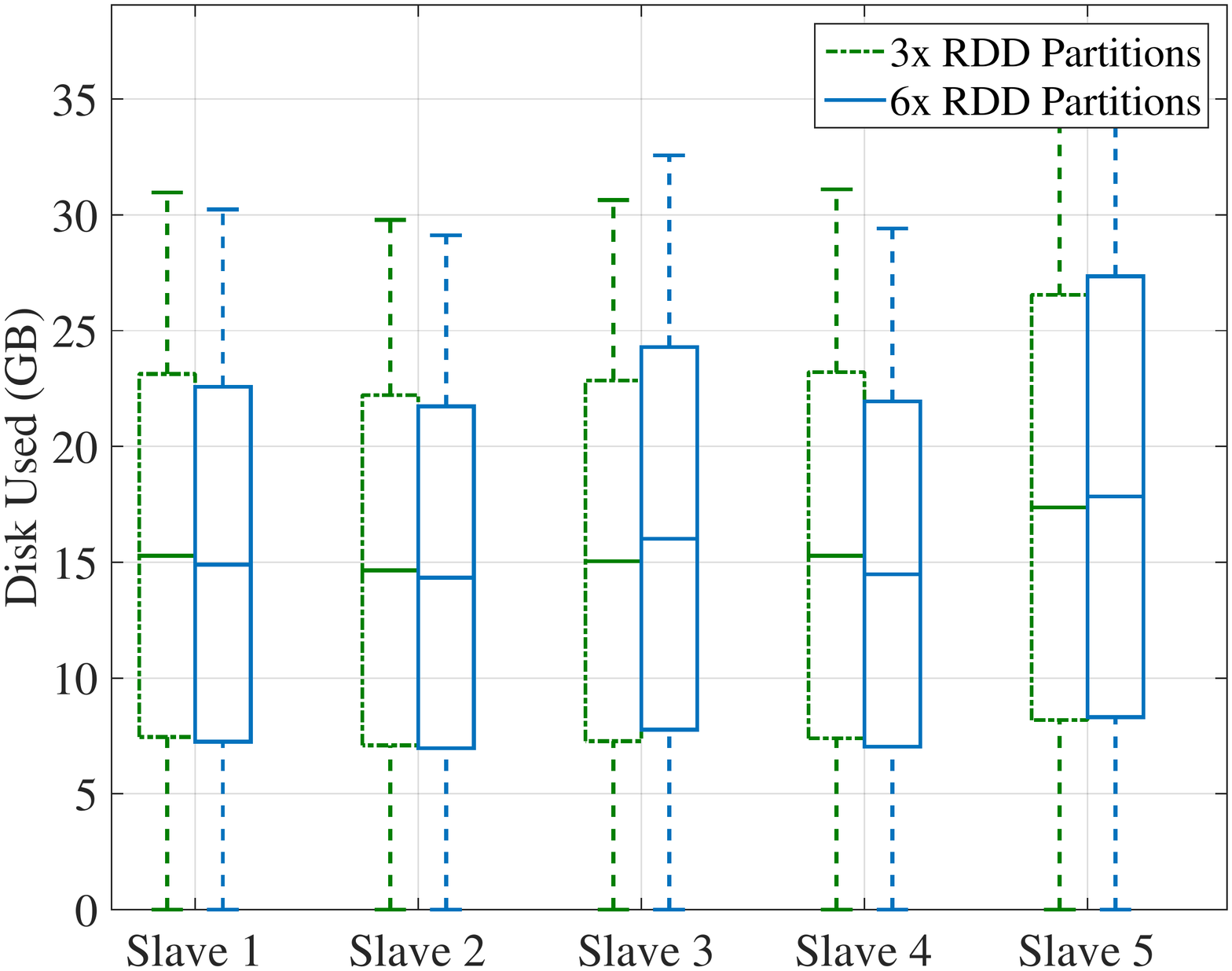}\\
\end{tabular}
\end{center}
\caption{Memory and disk usage per slave for the GS data: (a) memory usage when the memory-only model is applied, (b) disk usage when the memory-only model is applied, (c) memory usage when the memory-and-disk model is applied, (d) disk usage when the memory-and-disk model is applied.}\label{fig:cdl_gsmemory}
\end{figure*}

\noindent\textbf{Memory \& Disk Usage.} Figure~\ref{fig:cdl_hsmemory} and Figure~\ref{fig:cdl_gsmemory} present the memory usage per worker throughout the experiment duration for the HS data and GS data respectively, for $N=\{$3$x$, 6$x\}$. 

Considering the HS data (Fig.~\ref{fig:cdl_hsmemory}) the use of memory remains consistently at the maximum amount of memory allocated per worker for the computations of the assigned tasks. Similarly to the astrophysics use case (Section~\ref{psf_usecase}, Fig.~\ref{fig:psf_memory}), this result is aligned to the adopted persistence model (memory only) and the iterative nature of the optimization problem at hand, which entails subsequent increase of the respective number of computations over time. Nevertheless, opposed to the PSF use case, we observe similar memory usage in terms of statistical dispersion between the 25-th and 75-th percentiles, across all slaves involved. This is due to the current configuration of the cluster, which considers homogeneous resources allocation for all workers involved (i.e., 6 workers, each allocating 2.8GB RAM and 4 CPU cores). Finally, with regard to the impact of the number $N$ of partitions per RDD on the memory usage, the deviation on the memory usage is greater for $N=$3$x$ than the one recorder for $N=$6$x$. Indicatively, considering Slave 5 the dispersion of the memory usage equals to $\sim$60MB for $N=$3$x$, opposed to $\sim$30MB for $N=$6$x$. Similarly to the PSF use case, the smaller number of partitions ($N=$3$x$) results in fewer data blocks of greater size, thereby stimulating the persistence model to perform more on-the-fly calculations. By contrast, when $N=$6$x$, more data blocks of smaller size are generated, and subsequently, the persistence model become more relaxed. 

With regard to the case of the GS data (Fig.~\ref{fig:cdl_gsmemory}) interesting remarks can be derived depending on whether the memory-only or the memory-and-disk model is applied. Figures~\ref{fig:cdl_gsmemory}(a) and \ref{fig:cdl_gsmemory}(b) present the memory and disk usage respectively for the memory-only model, while Fig.~\ref{fig:cdl_gsmemory}(c)-(d) illustrate the memory and disk usage for the memory-and-disk model. Similarly to the case of HS data and the astrophysics use case, the adoption of the memory-only model results in employing the maximum amount of memory per worker for the computations of the assigned tasks, while the smaller the value of $N$, the more stable the behaviour of each worker. Nevertheless, opposed to the case of HS data and the astrophysics use case, the increased size of the problem ($P=$17$\times$17, $M=$9$\times$9) result into using disk space for storing intermediate results not fitting into the memory(Fig.~\ref{fig:cdl_gsmemory}(b)). When the memory-and-disk model is applied we observe that the use of memory decreases to less than 500MB on all slaves, while the value of the number of partitions $N$ has no essential impact on the statistical variation of the memory usage. Specifically, the dispersion between the 25-th and 75-th percentile remains at $\sim$160MB for all slaves and different values of $N$. The decreased use of memory is accompanied by a substantial increase in the disk usage, as illustrated in Fig.~\ref{fig:cdl_gsmemory}(c)-\ref{fig:cdl_gsmemory}(d); the median disk volume increases from $\sim$5GB to $\sim$15GB when the cluster configuration transits from the memory-only to the memory-and-disk model. 

\begin{figure*}[ht!]
\setlength{\abovecaptionskip}{0.5pt} 
\begin{center}
\begin{tabular}{c}
\small{(a)}\\
\includegraphics[scale=0.3]{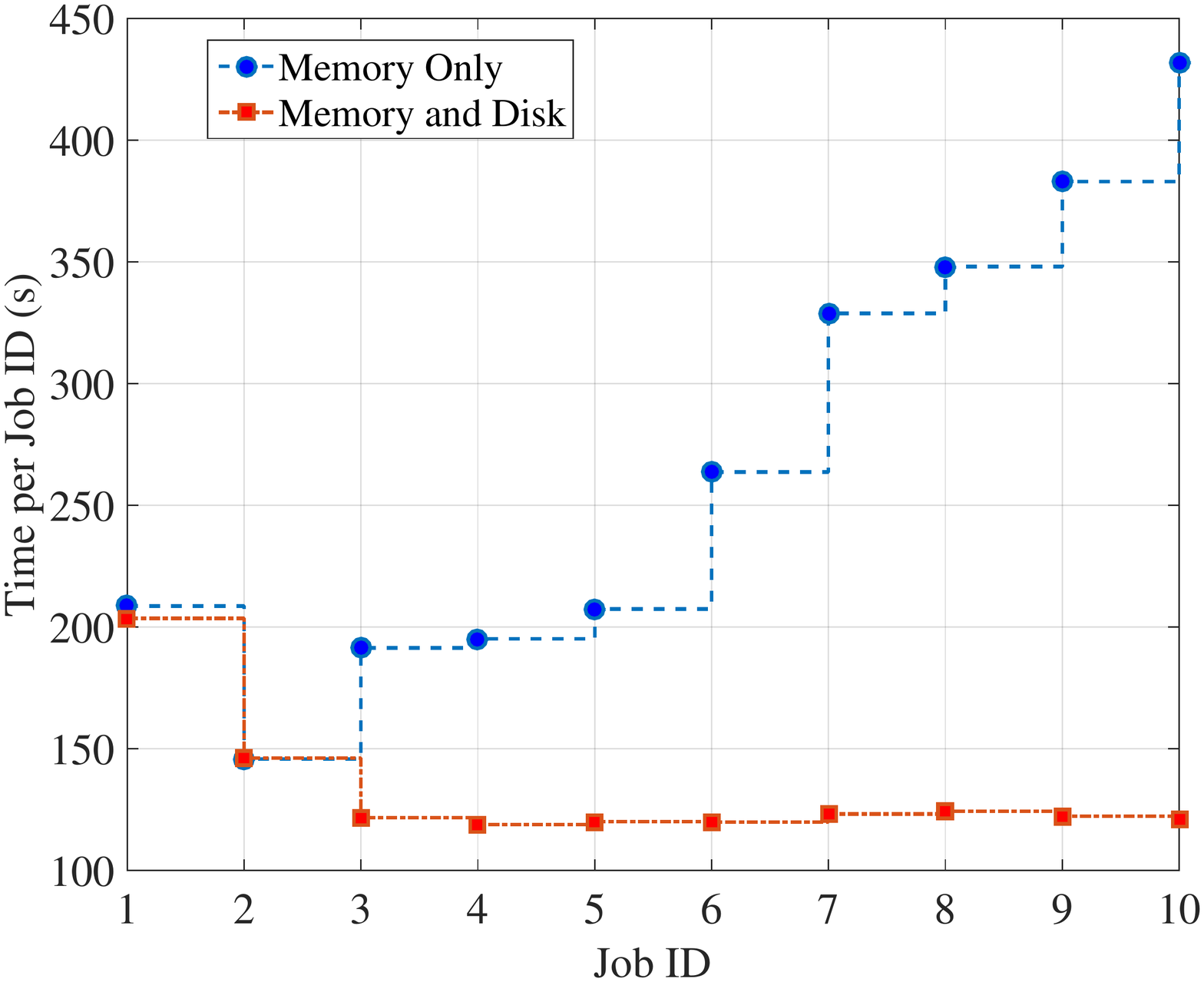}\\
\begin{tabular}{cc}
\small{(b)} & \small{(c)}\\
\includegraphics[scale=0.28]{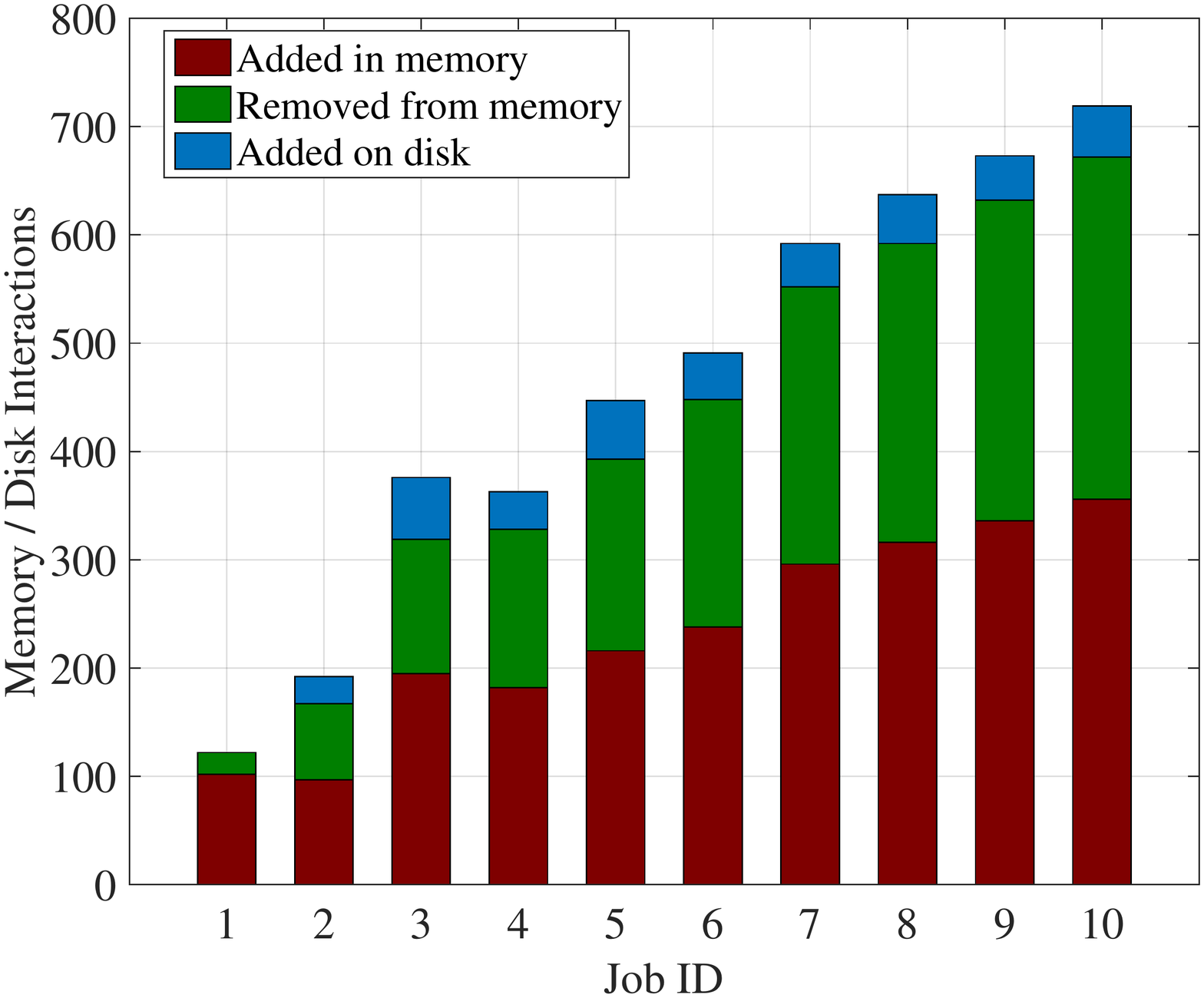} & \includegraphics[scale=0.28]{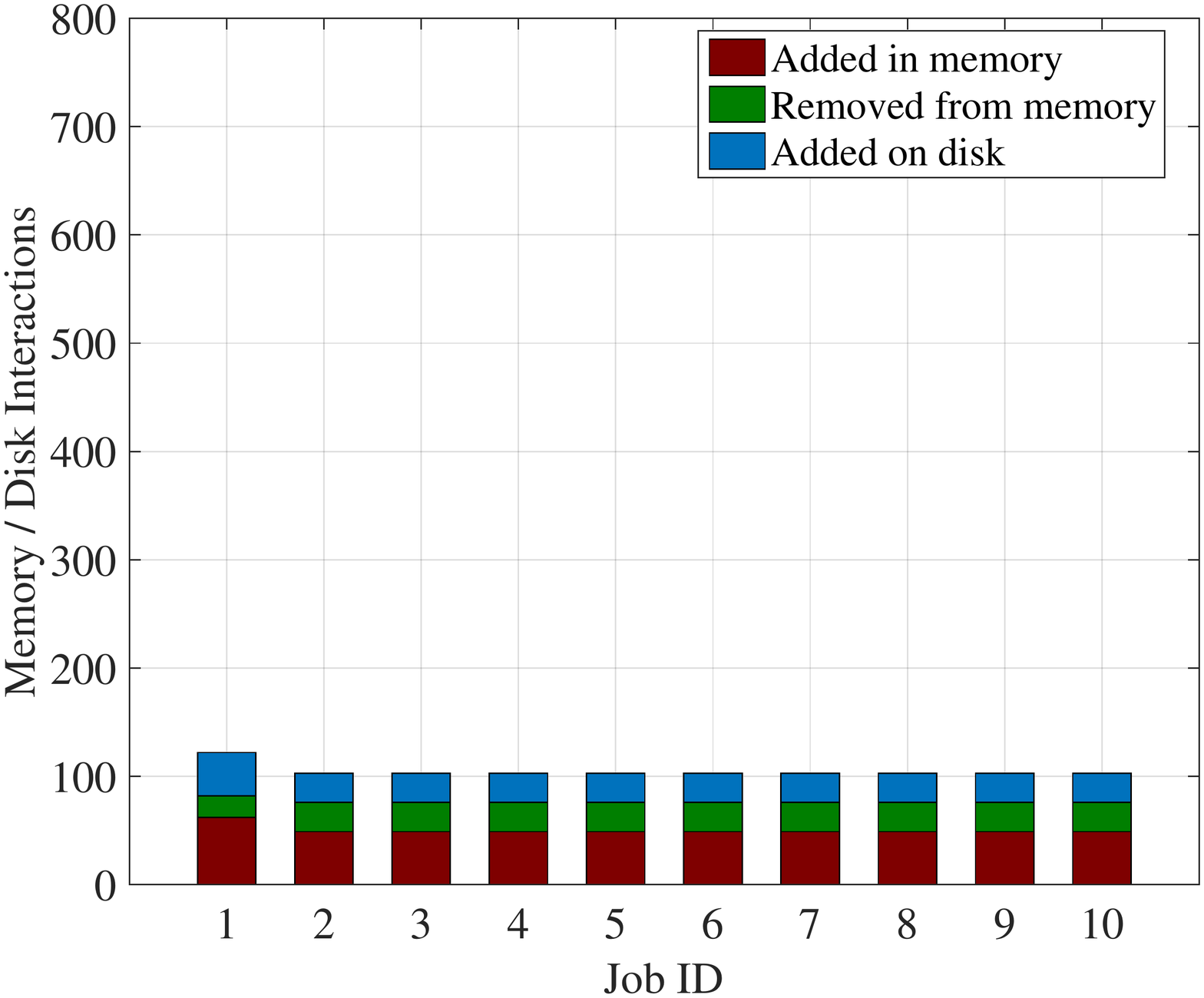}\\
\end{tabular}
\end{tabular}
\end{center}
\caption{Time execution and memory-disk interactions for the GS data: (a) time execution on Slave 1 for 10 subsequent Spark jobs, (b) memory and disk interactions when the memory-only model is applied, (c) memory and disk interactions when the memory-and-disk model is applied.}\label{fig:cdl_memdisk}
\end{figure*}

The increased disk usage is consistent to the persistence model and in order to investigate how the disk overhead affects the time performance, we present in Fig.~\ref{fig:cdl_memdisk} the time execution (Fig.~\ref{fig:cdl_memdisk}(a)), along with the memory and disk interactions when either the memory-only (Fig.~\ref{fig:cdl_memdisk}(b)) or the memory-and-disk  (Fig.~\ref{fig:cdl_memdisk}(c)) model is applied, over ten subsequent Spark tasks. The results indicate that \emph{the use of memory-only model is not advisable for this scenario}; the time for the execution of the subsequent iterations for calculating $\mathbf{X}_{h}^{*}$, and $\mathbf{X}_{l}^{*}$ increases over time. Specifically, the time needed for completing 10 subsequent Spark jobs increases 200 sec$\rightarrow$430 sec when the memory-only model is applied. By contrast, when the memory-and-disk model is applied, the execution of the same 10 subsequent jobs reduces 200 sec$\rightarrow$120 sec. This behaviour relates to the needed memory interactions (add, remove, add on disk) that take place depending on the persistence model. As shown in Fig.~\ref{fig:cdl_memdisk}(b), the memory-only model, which entails the on-demand recalculation of intermediate results, imposes an increasing trend of adding-removing results from memory, in order to free resources for the execution of subsequent iterations. On the other hand, when the memory-and-disk model is applied (Fig.~\ref{fig:cdl_memdisk}(c)) the number of add-remove memory interactions remains consistent throughout the execution of the Spark jobs, since intermediate results are directly stored on the disk instead of recalculating them when they are needed. 

\begin{figure*}[h!]
\setlength{\abovecaptionskip}{0.5pt} 
\begin{center}
\begin{tabular}{cc}
\small{(a)} & \small{(b)}\\
\includegraphics[scale=0.28]{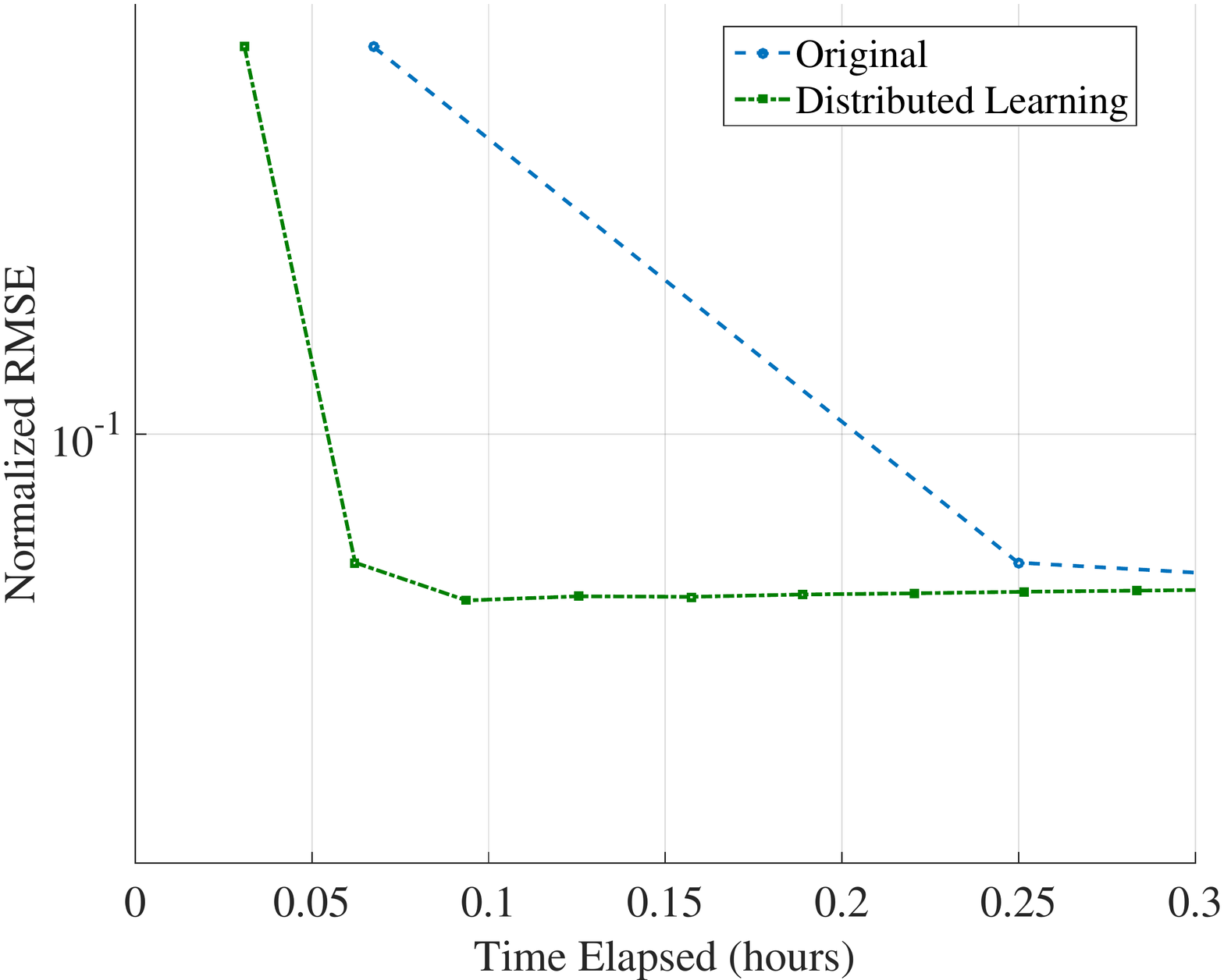} & \includegraphics[scale=0.28]{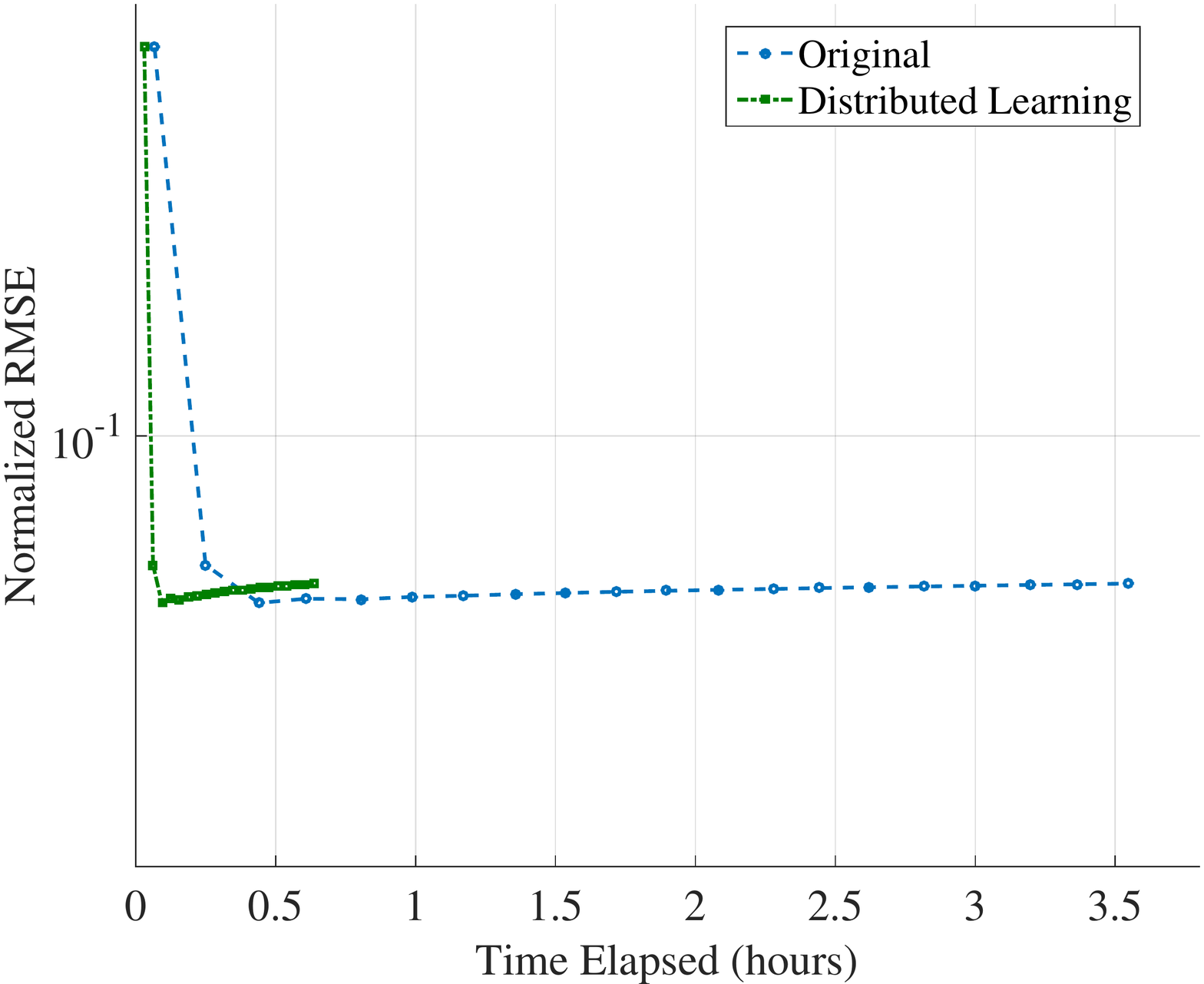}\\
\small{(c)} & \small{(d)}\\
\includegraphics[scale=0.28]{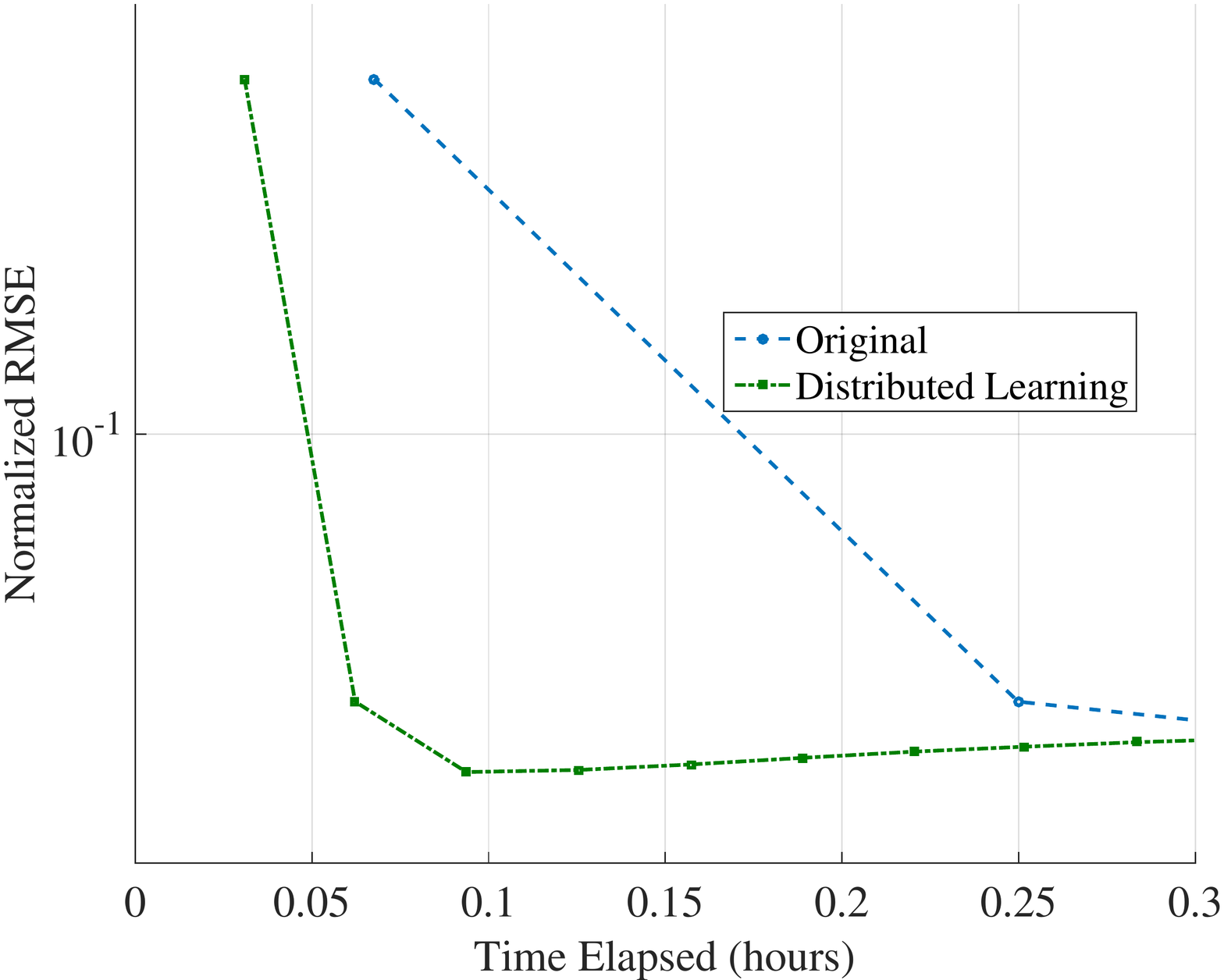} & \includegraphics[scale=0.28]{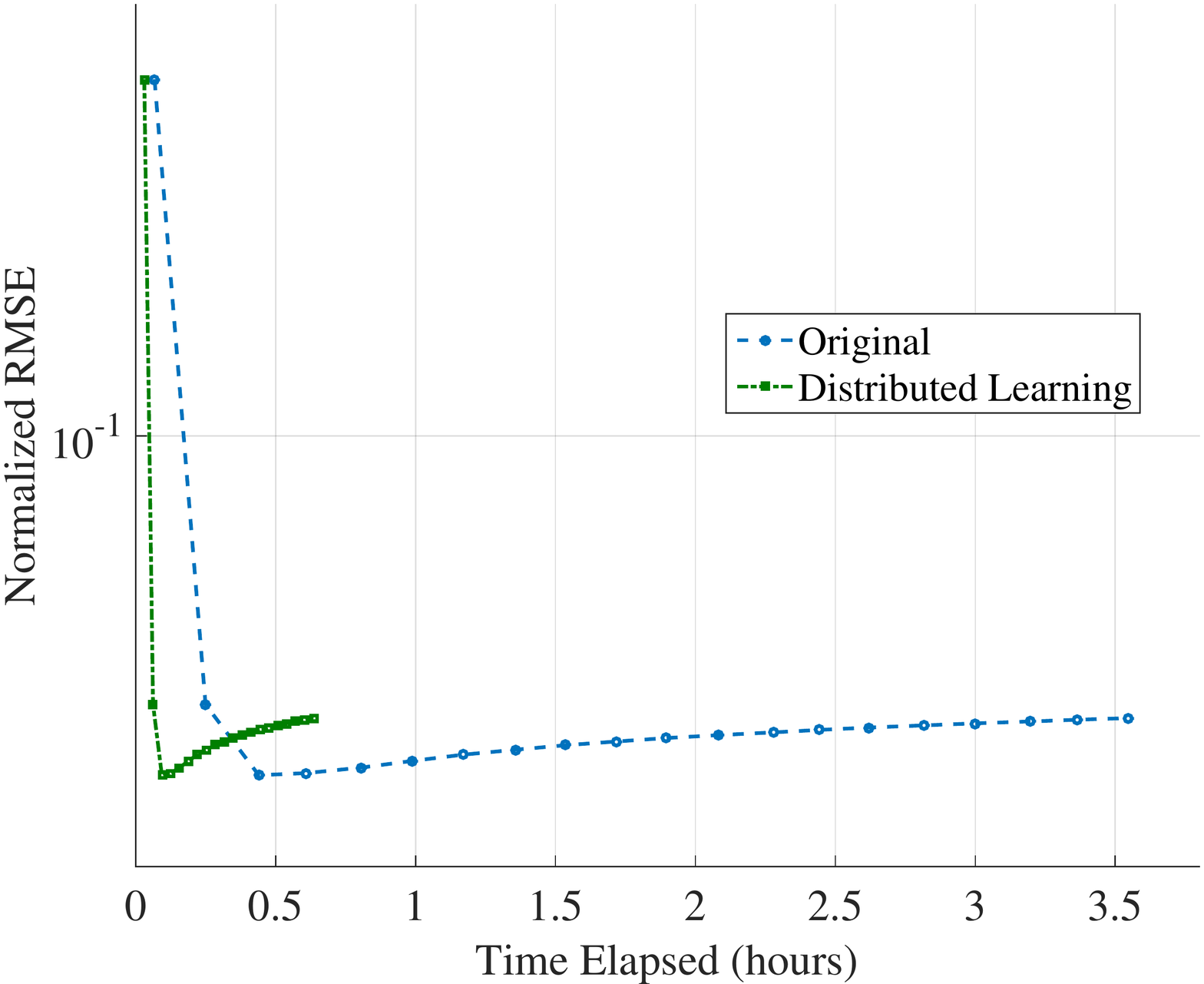}\\
\end{tabular}
\end{center}
\caption{The reconstruction error between calculated dictionaries and augmented Lagrangian function for the GS data and memory-and-disk persistence model: (a) high resolution during the first 20 minutes of the experiment, (b) high resolution, total duration of the experiment, (c) low resolution during the 20 minutes of the experiment, (d) low resolution total duration of the experiment.}\label{fig:cdl_convergence}
\end{figure*}
\noindent\textbf{Convergence Rate.} Ultimately, Fig.~\ref{fig:cdl_convergence} illustrates the convergence behavior of reconstruction error between the calculated dictionaries ($\mathbf{X}_{h}^{*}$, $\mathbf{X}_{l}^{*}$) and the augmented Lagrangian function versus the time elapsed, when either the sequential or the distributed SCDL approach (Algorithm~\ref{scdl_parallel}, $N=$3$x$) is adopted for the GS data, considering the memory-and-disk persistence model. For both high (Fig.~\ref{fig:cdl_convergence}(a)) and low (Fig.~\ref{fig:cdl_convergence}(c)) resolution dictionaries, the reconstruction error starts converging within the first 20 minutes of experiment when the proposed architecture is adopted. This is in sharp contrast to the sequential approach which can only complete two iterations within the same time period. Overall, the distributed learning approach is 65.7\% faster than the conventional one; the completion time of the standalone approach equals to $\sim$3.5 hours, opposed to the parallelized version, which does not exceed 0.8 hours (Fig.~\ref{fig:cdl_convergence}(b), Fig.~\ref{fig:cdl_convergence}(d)). These results highlight that the herein proposed solution is extremely beneficial in terms of time response for enabling large-scale joint dictionary training schemes.

\subsection{Discussion and Practical Guidelines}
The evaluation studies indicate the effectiveness of the proposed architecture on jointly processing multiple imaging datasets. Even so, each application scenario requires a different approach in order to be treated as a distributed learning problem. Profiling both the problem and the data characteristics in order to compensate the performance bottlenecks is considered a critical aspect for providing an alternative, distributed solution that outperforms conventional approaches. Comprehensively, the key findings derived by this study can be summarized as follows:
\begin{itemize}
\item While being a generic purpose distributed computing framework, the core module of Spark herein adopted yields extremely promising results for solving image processing problems at scale. As highlighted in Fig.~\ref{fig:psf_scalability} and Fig.~\ref{fig:cdl_scalability} for the astrophysics and remote sensing use cases respectively, our architecture offers a scalable solution, providing more than 50\% improvement in time response, when the number of available cores becomes 6 times greater than the original approach. At the same time, the convergence behavior of the optimization process becomes substantially faster compared to the sequential counterpart (75\% for the astrophysics use case, 65.7\% for the remote sensing use case).
\item With regard to the cluster behavior, the scale of an input imaging dataset is relevant to the capacity of the cluster, and the demands of the solving approach; the overhead introduced by Spark for the distributed calculations may hinder the overall performance, when the dataset is relatively small compared to the computational capacity of the cluster. In such cases it is preferable to retain a small number of partitions $N$, in order to avoid introducing unnecessary shuffling overhead.
\item On a similar note, the benefit of having an increased number of partitions is related to the memory usage per worker. Results on both use cases (Fig.~\ref{fig:psf_memory} and Fig~\ref{fig:cdl_hsmemory}-\ref{fig:cdl_gsmemory}) point out the trade off between the RAM availability and the size of the input problem; When the memory per worker is limited with respect to size of the input data it is considered preferable to increase the number of partitions of the RDD. It may increase the time response, due to the increased number of tasks, however it will yield a more reliable solution within the lifetime of the program execution.
\item The cluster configuration in terms of homogeneity of computational resources can also affect the memory behavior. Providing workers with similar RAM size and number of computational cores results in smaller deviation on the memory usage across all slaves involved. This is evident when comparing the memory behavior results (Fig.~\ref{fig:psf_memory} versus Fig.~\ref{fig:cdl_hsmemory}-\ref{fig:cdl_gsmemory}) when either the astrophysics (five slaves$\rightarrow$five workers, wherein one worker provides double number of CPU cores than the remaining four) or the remote sensing (five slaves$\rightarrow$six workers, all workers provide equivalent number of CPU cores) problem is considered. 
\item As highlighted in the super-resolution use case, the impact of the persistence model is crucial when the memory per worker is limited and the use of disk space is unavoidable. In such cases, the memory-and-disk model, responsible for storing intermediate results on disk, is preferable; it eliminates the necessity of adding-removing results from memory, and thereby improve the time performance of subsequent calculations (e.g. Fig.~\ref{fig:cdl_memdisk}).
\end{itemize}

\section{Conclusions}\label{sec:conclusions}
In this work we propose an Spark-compliant architecture for performing distributed learning over bundled scientific imaging data. We present the respective algorithm parallelization for two high-impact use cases namely: (a) the computational astrophysics domain (space-variant deconvolution of noisy galaxy images), (b) the remote sensing domain (super-resolution using sparse coupled dictionary training). The respective software libraries and datasets are publicly available at~\cite{papercode}. The evaluation studies highlight the practical benefits of changing the implementation exemplar and moving towards distributed computational approaches; while employing commodity hardware the results for both application scenarios indicate an improvement greater than 60\% in time response terms against the conventional computing solutions. In addition, the trade off between memory availability and computational capacity has been revealed, while the offered insights draft the roadmap for further improvements on distributed scientific imaging frameworks.

\section*{Acknowledgment}
This work was funded by the DEDALE project (no. 665044) within the H2020 Framework Program of the European Commission.

\section*{References}
\bibliography{refs}
%

\end{document}